\documentclass[twocolumn,preprintnumbers,amsmath,amssymb,prb,superscriptaddress]{revtex4}
\usepackage[latin1]{inputenc}
\usepackage{dcolumn,color,dsfont,bm,graphicx,bbm}
\usepackage[english]{babel}
\usepackage[bbgreekl]{mathbbol}
\DeclareMathOperator{\Tr}{Tr}
\DeclareMathOperator{\Rea}{Re}
\DeclareMathOperator{\Ima}{Im}

\def\bbm[#1]{\mbox{\boldmath $#1$}}

\makeatletter
\def\amsbb{\use@mathgroup \M@U \symAMSb}
\makeatother

\usepackage{hyperref}
\hypersetup{colorlinks,
linkcolor=black,
filecolor=green,
urlcolor=blue,
citecolor=black}

\graphicspath{{./Figures/}}

\begin{document}

\title{Surface-mode-assisted amplification of radiative heat transfer between nanoparticles}

\author{Riccardo Messina}
\email{riccardo.messina@institutoptique.fr}
\affiliation{Laboratoire Charles Fabry, UMR 8501, Institut d'Optique, CNRS, Universit\'{e} Paris-Saclay, 2 Avenue Augustin Fresnel, 91127 Palaiseau Cedex, France}

\author{Svend-Age Biehs}
\email{s.age.biehs@uni-oldenburg.de}
\affiliation{Institut f\"{u}r Physik, Carl von Ossietzky Universit\"{a}t, D-26111 Oldenburg, Germany}

\author{Philippe Ben-Abdallah}
\email{pba@institutoptique.fr}
\affiliation{Laboratoire Charles Fabry, UMR 8501, Institut d'Optique, CNRS, Universit\'{e} Paris-Saclay, 2 Avenue Augustin Fresnel, 91127 Palaiseau Cedex, France}

\date{\today}

\begin{abstract}
We show that the radiative heat flux between two nanoparticles can be significantly amplified when they are placed in proximity of a planar substrate supporting a surface resonance. The amplification factor goes beyond two orders of magnitude in the case of dielectric nanoparticles, whereas it is lower in the case of metallic nanoparticles. We analyze how this effect depends on the frequency and on the particles-surface distance, by clearly identifying the signature of the surface mode producing the amplification. Finally, we show how the presence of a graphene sheet on top of the substrate can modify the effect, by making an amplification of two orders of magnitude possible also in the case of metallic nanoparticles. This long range amplification effect should play an important role in the thermal relaxation dynamics of nanoparticle networks.
\end{abstract}

\maketitle

\section{Introduction}

Two bodies at different temperatures placed in vacuum experience an energy exchange mediated by photons. The Stefan-Boltzmann law states that this radiative heat transfer is limited in the far field, i.e. for distances larger than the thermal wavelength $\lambda_T=\hbar c/k_BT$ (close to 8\,$\mu$m at room temperature), by the amount exchanged between two blackbodies, defined as ideal bodies absorbing all incoming radiation. The pioneering works of Rytov~\cite{Rytov}, Polder and van Hove~\cite{PoldervH} showed that in the opposite regime, the near field, this limitation does not hold and the flux can surpass even by several orders of magnitude the blackbody limit. This amplification can be dramatic when the bodies support surface modes~\cite{LoomisPRB94,PendryJPhysCondensMatter99,VolokitinPRB01,VolokitinPRB04,JoulainSurfSciRep05,VolokitinRevModPhys07}. More specifically, this amplification typically happens for dielectrics, having resonance frequencies lying in the infrared range, whereas the surface modes of metals (typically in the ultraviolet range) do not participate to the effect for temperatures close to the ambient temperature. Several experiments on near-field heat transfer have been realized in different geometries so far, establishing a quite solid agreement between measurements and theory~\cite{KittelPRL05,HuApplPhysLett08,NarayanaswamyPRB08,RousseauNaturePhoton09,ShenNanoLetters09,KralikRevSciInstrum11,
OttensPRL11,vanZwolPRL12a,vanZwolPRL12b,KralikPRL12,KimNature15,SongNatureNano15,StGelaisNatureNano16,KloppstecharXiv,WatjenAPL16}.

A remarkable theoretical effort in this domain has been devoted to the study of the heat exchange, both in the stationary and dynamical regimes, between two or more nanoparticles~\cite{PBA-APL2006,PBA-PRB2008,POC2008,PBA-PRL2011,TschikinEurPhysJB12,YannopapasPRL13,YannopapasJPhysChemC13,BiehsAgarwal2013,BaffouOptExpress09,BaffouPRB10,BaffouLaserPhotonicsRev13,Add1,Add2,Add3,Add4,Add5,Add6,Add7,Add8,Add9,Add10,Add11,Add12}. The reduced size of the particles enables to perform the dipole approximation, in which each particle is assumed to be as a pointlike source and its interaction with the field described in terms of a dielectric \textcolor{black}{and/or magnetic} dipole. \textcolor{black}{This assumption simplifies} considerably the calculations but \textcolor{black}{limits} the validity of the results to distances larger than the typical size of the nanoparticles. Generally speaking, previous works have been focused on the active control of the cooling and heating of nanoparticles, either in vacuum or in proximity of an interface, as well as of the temperature profile within a collection of nanoparticles.

We focus here on a specific aspect of the radiative heat transfer between two nanoparticles. Guided by the major role played by surface modes on the value and spectral properties of near-field heat transfer, we study how the proximity of two nanoparticles to a substrate supporting such a mode can amplify the heat flux exchanged between them.
In particular, we focus on the case of a dielectric (silicon carbide) substrate, and consider both scenarios of dielectric and metallic nanoparticles on top of it. We predict a flux amplification which goes beyond two orders of magnitude in the case of dielectric nanoparticles, while the enhancement is close to \textcolor{black}{6} for metallic nanoparticles. The physics behind this effect is studied both in terms of spectral properties and with respect to the distance between particles and substrate, in order to well identify the role played by the surface mode. Finally, we also address the effect of a graphene sheet placed on top of the substrate, showing that it dramatically increases the effect in the case of metallic \textcolor{black}{nanoparticles}.

The paper is structured as follows. In Sec.~\ref{SecGreen}, we introduce the geometry of our system, define the Green function in the absence and presence of substrate, and give the expression of the heat flux between the two nanoparticles. In Sec.~\ref{SecAmpli}, we present our main results concerning the surface-mode-induced amplification of heat flux. Section \ref{SecGraphene} is dedicated to the role of a graphene sheet on top of the substrate. We finally in Sec.~\ref{SecConclusions} give some conclusive remarks and perspectives.

\section{Green function and heat transfer between nanoparticles}\label{SecGreen}

Let us consider two nanoparticles having coordinates $\mathbf{R}_1$ and $\mathbf{R}_2$ respectively, located close to the plane $z = 0$, separating vacuum ($z>0$) from a region ($z<0$) occupied by a non-magnetic medium having dispersive electric permittivity $\varepsilon(\omega)$. We assume that the two particles are placed at the same distance $z$ from the interface, while $d$ is the distance between them. In virtue of the rotational symmetry of the system with respect to the $z$ axis, we can choose, without loss of generality, the coordinates of the two particles to be $\mathbf{R}_1 = (0,0,z)$, $\mathbf{R}_2 = (d,0,z)$.

In the following we will calculate the radiative heat transfer between the two nanoparticles. More specifically, we want to investigate how the presence of the substrate, and in particular of a surface mode existing at the interface with vacuum, is able to modify and possibly amplify this energy-exchange mechanism. For the sake of simplicity, we will work in the framework of the dipolar approximation, according to which the two particles are described as pointlike \textcolor{black}{sources}. This assumption is valid as long as the length scales involved in the system are large compared to the size of the nanoparticles. \textcolor{black}{We will assume that the two nanoparticles are identical spheres of radius $R = 5\,\text{nm}$, and limit our calculations to values of the particle-surface distance $z$ and particle-particle distance $d$ both larger than 50\,nm. This implies that the particule radius satisfies $R\gg d,z,\lambda$ ($\lambda$ being a relevant wavelength participating to the energy exchange), guaranteeing the validity of the dipolar approximation. The description of the optical response of a pointlike nanoparticle is easily made in terms of a series development of the Mie coefficients, describing the scattering on a sphere~\cite{BH}. According to the nature (dielectric or metallic) of the nanoparticles, the relevant terms can be the electric and magnetic frequency-dependent polarizabilities $\alpha_E^{(0)}(\omega)$ and $\alpha_H^{(0)}(\omega)$. In the limit $R\ll \delta$ ($\delta$ being the skin depth of the given material), these can be written under the well-known Clausius-Mossoti form
\begin{equation}\label{alphas}\begin{split}
  \alpha_E^{(0)}(\omega) &= 4\pi R^3\frac{\varepsilon(\omega)-1}{\varepsilon(\omega)+2},\\
  \alpha_H^{(0)}(\omega) &= \frac{2\pi}{15} R^3\Bigl(\frac{\omega R}{c}\Bigr)^2[\varepsilon(\omega)-1],\\
\end{split}\end{equation}
$R$ and $\varepsilon(\omega)$ being the radius and the electric permittivity of the particle, respectively. To any of the two polarizabilities} we apply the radiative correction, discussed e.g in Refs.~\onlinecite{CarminatiOptCommun06,AlbaladejoOptExpress10}, thus obtaining the dressed polarizability
\begin{equation}
    \alpha(\omega)=\frac{\alpha^{(0)}(\omega)}{1-i\frac{k_0^3}{6\pi}\alpha^{(0)}(\omega)},
\end{equation}
where $k_0=\omega/c$. We finally need to introduce the \textcolor{black}{modified polarizability}
\begin{equation}
  \chi(\omega) =\Ima[\alpha(\omega)]-\frac{k_0^3}{6\pi}|\alpha(\omega)|^2
\end{equation}
which appears in the fluctuation-dissipation theorem describing dipole fluctuations and avoids unphysical effects (for a more detailed discussion see Ref.~\onlinecite{ManjavacasPRB12}).

\textcolor{black}{We will assume that the system is thermalized at $T=300\,$K and that the temperature of one of the two nanoparticles is slightly increased to $300\,\text{K}+\Delta T$. This generates a non-vanishing heat flux $\varphi$ on the other nanoparticle, which can be entirely ascribed to an energy exchange between the two nanoparticles. The ratio between the flux $\varphi$ and the temperature difference $\Delta T$ defines, in the limit $\Delta T\to 0$, the conductance $G$. This is the quantity we are going to calculate in the following, addressing in particular the question of how $G$ is modified by the presence of the substrate. Thus, in our calculation the substrate purely acts as a boundary condition modifying the way in which the direct exchange between the particles takes place.} The conductance $G$ in the case of two identical particles can be conveniently expressed in terms on the Green function describing the system as~\cite{PBA-PRL2011}
\begin{equation}\label{ExprG}
    G = 4\int_0^{+\infty}\!\!\frac{d\omega}{2\pi}\,\hbar\omega\,k_0^4\,n'(\omega,T)\chi^2\Tr\bigl(\mathcal{G}\mathcal{G}^\dag\bigr)
\end{equation}
where $\mathcal{G}$ denotes the dyadic Green tensor of the full system which \textcolor{black}{is written} in terms of Green tensor $\mathds{G}$ of a single interface as
\begin{equation}\label{fullGreen}
    \mathcal{G} =\mathds{M}^{-1}\mathds{G},
\end{equation}
$\mathds{M} = \mathds{1}-k_0^4\alpha_1\alpha_2\mathds{G}\mathds{G}^\text{T}$ representing the \textcolor{black}{multiple reflections} between the two particles. In expression (\ref{ExprG}) the frequency dependence of material-depending quantities ($\chi$ and $\mathcal{G}$) has been omitted for simplicity reasons and $n'(\omega,T)$ denotes the derivative with respect to $T$ of the Bose--Einstein distribution
\begin{equation}
   n(\omega,T) = \biggl[\exp\biggl(\frac{\hbar\omega}{k_B T}\biggr) - 1\biggr]^{-1}.
\end{equation}
In the infrared range the multireflections can be neglected  (see Ref.~\onlinecite{MessinaPRB13}) so that in this spectral range the equality $\mathcal{G} =\mathds{G}$ holds. \textcolor{black}{We remark here that for a particle described in terms of an electric dipole the modified polarizability $\chi$ appearing in Eq.~\eqref{ExprG} is the one derived from the electric polarizability $\alpha_E^{(0)}$ and the Green function is the electric-electric one $\mathds{G}_{EE}$. On the contrary, the magnetic contribution to the conductance is obtained by using the modified polarizability $\chi$ derived from the magnetic polarizability $\alpha_H^{(0)}$ and the magnetic-magnetic Green function $\mathds{G}_{HH}$.}

In the presence of a vacuum-material interface, the Green function can be written as
\begin{equation}\label{Gsum}
 \mathds{G} = \mathds{G}^{(0)} + \mathds{G}^{(\text{sc})},
\end{equation}
i.e. separated into a vacuum contribution and a scattering part which depends on the interface reflection coefficients and goes to zero in the absence of the interface. The vacuum contribution to the Green function reads
\textcolor{black}{\begin{equation}\label{G0}
 \mathds{G}_{EE}^{(0)} = \mathds{G}_{HH}^{(0)} = \frac{e^{ik_0d}}{4\pi k_0^2d^3}\begin{pmatrix}a & 0 & 0\\0 & b & 0\\0 & 0 & b\end{pmatrix},\\
\end{equation}}
where $a = 2 - 2 i k_0 d$, $b = k_0^2 d^2 + i k_0 d - 1$.

The scattering contribution to the \textcolor{black}{electric-electric} Green's function can be written as an integral with respect to the modulus $k=|\mathbf{k}|$ of the wavevector $\mathbf{k}=(k_x,k_y)$ on the $x-y$ plane as follows~\cite{Novotny-book}
\textcolor{black}{\begin{equation}\label{Gsc}
 \mathds{G}_{EE}^{(\text{sc})} = \int_0^{+\infty} \frac{dk}{2\pi}\,\frac{ike^{2ik_zz}}{2k_0^2k_z}\bigl(r_s\mathbb{S} + r_p\mathbb{P}\bigr).
\end{equation}}
In this expression $k_z = \sqrt{k_0^2 - k^2}$ is the $z$ component of the wavevector in vacuum, while $r_s$ and $r_p$ are the ordinary Fresnel coefficients associated with the two polarizations. Defining as $k_{zm} = \sqrt{\varepsilon(\omega)k_0^2 - k^2}$ the $z$ component of the wavevector inside the material, these coefficients are given by
\begin{equation}
 r_s = \frac{k_z - k_{zm}}{k_z + k_{zm}},\quad r_p = \frac{\varepsilon(\omega) k_z - k_{zm}}{\varepsilon(\omega) k_z + k_{zm}}.
\end{equation}
Finally, the matrices $\mathbb{S}$ and $\mathbb{P}$ are defined as
\begin{equation}\begin{split}
\mathbb{S} &= \begin{pmatrix}k_0^2A^+ & 0 & 0\\0 & k_0^2A^- & 0\\0 & 0 & 0\end{pmatrix},\\
\mathbb{P} &= \begin{pmatrix}-k_z^2A^- & 0 & - kk_zB_1\\0 & -k_z^2A^+ & 0\\kk_zB_1 & 0 & k^2B_0\end{pmatrix},
\end{split}\end{equation}
being
\begin{equation}
 A^\pm = \frac{1}{2}\Bigl[J_0(kd) \pm J_2(kd)\Bigr],\hspace{0.1cm}B_n = i^n J_n(kd),
\end{equation}
where $J_n$ is the cylindrical Bessel function of order $n$. \textcolor{black}{The scattering part of the magnetic-magnetic Green function can be easily obtained from $\mathds{G}_{EE}^{(\text{sc})}$ by exchanging $r_s$ and $r_p$ in Eq.~\eqref{Gsc}.}

Based on Eq.~\eqref{Gsum}, we remark that
\begin{equation}\begin{split}
 \Tr\bigl(\mathds{G}\mathds{G}^\dag\bigr) &= \Tr\bigl(\mathds{G}^{(0)}\mathds{G}^{(0)\dag}\bigr) + \Tr\bigl(\mathds{G}^{(\text{sc})}\mathds{G}^{(\text{sc})\dag}\bigr)\\
 &\,+ 2\Rea\bigl[\Tr\bigl(\mathds{G}^{(0)}\mathds{G}^{(\text{sc})\dag}\bigr)\bigr],
\end{split}\end{equation}
allowing us to decompose the conductance as
\begin{equation}\label{ContributionsG}
 G = G^{(0,0)} + G^{(\text{sc},\text{sc})} + G^{(0,\text{sc})},
\end{equation}
i.e. as the sum of the vacuum contribution and two further contributions associated with the presence of an interface, more specifically a scattering and a crossed term, $G^{(\text{sc},\text{sc})}$ and $G^{(0,\text{sc})}$ respectively. These contributions will be discussed in the following together with the total conductance $G$.

\section{Surface-mode amplification of heat flux}\label{SecAmpli}

\begin{figure}[t!]
\includegraphics[width=0.5\textwidth]{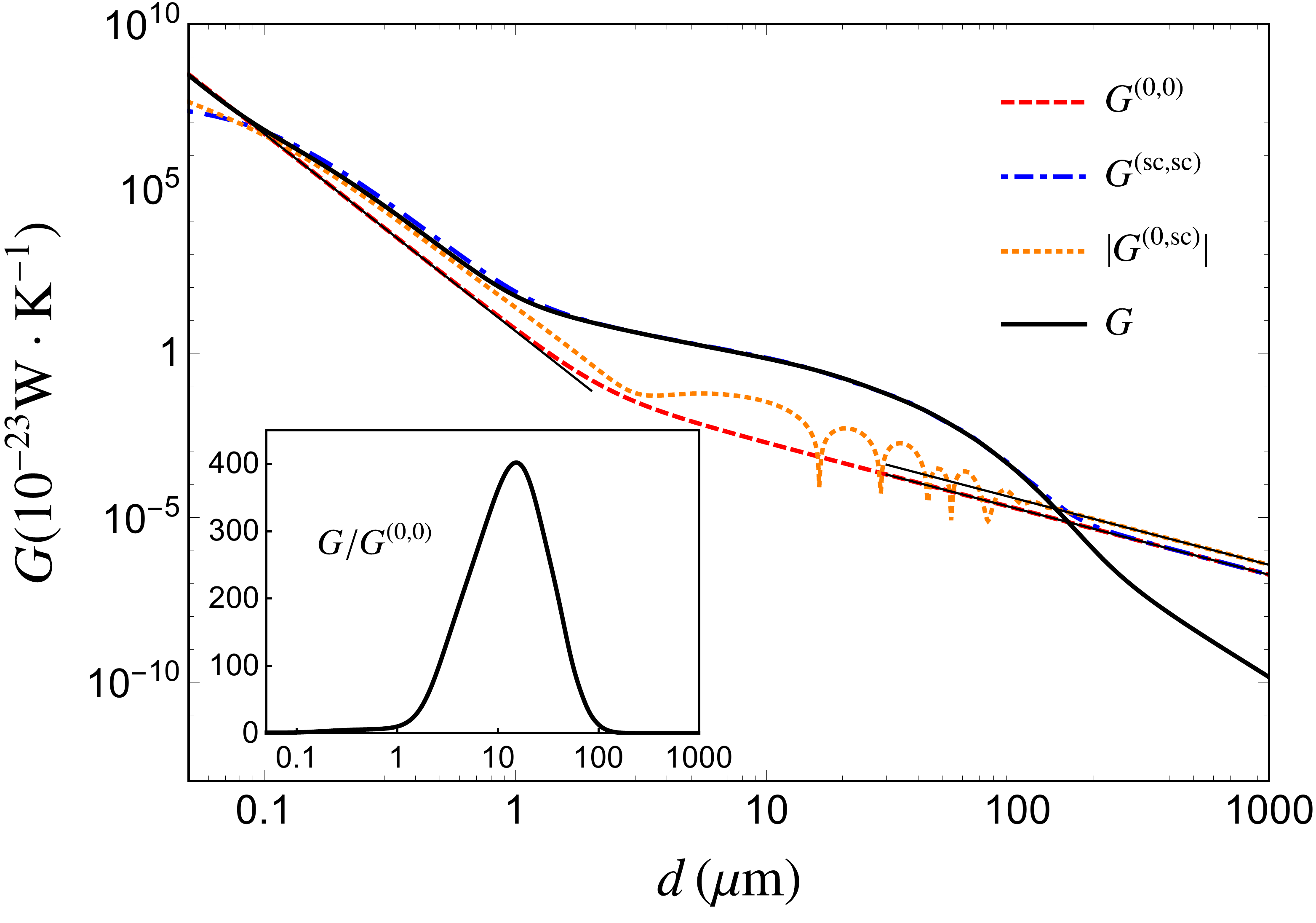}
\caption{Total conductance $G$ (black solid line) and single contributions $G^{(0,0)}$ (red dashed line), $G^{(\text{sc,sc})}$ (blue dot-dashed line) and $G^{(\text{0,sc})}$ (in absolute value, orange dotted line) as defined in Eq.~\eqref{ContributionsG} between two SiC nanoparticles at distance $d$, placed at distance \textcolor{black}{$z=50\,$nm} from a SiC substrate. The thin black lines correspond to the \textcolor{black}{small}- long-distance asymptotic behaviors of $G^{(0,0)}$ given in Eq.~\eqref{d-6-2}. The inset shows the ratio between conductances in the presence and absence of substrate as a function of $d$.}
\label{SiC_01}
\end{figure}

\begin{figure*}[t!]
\includegraphics[height=0.235\textwidth]{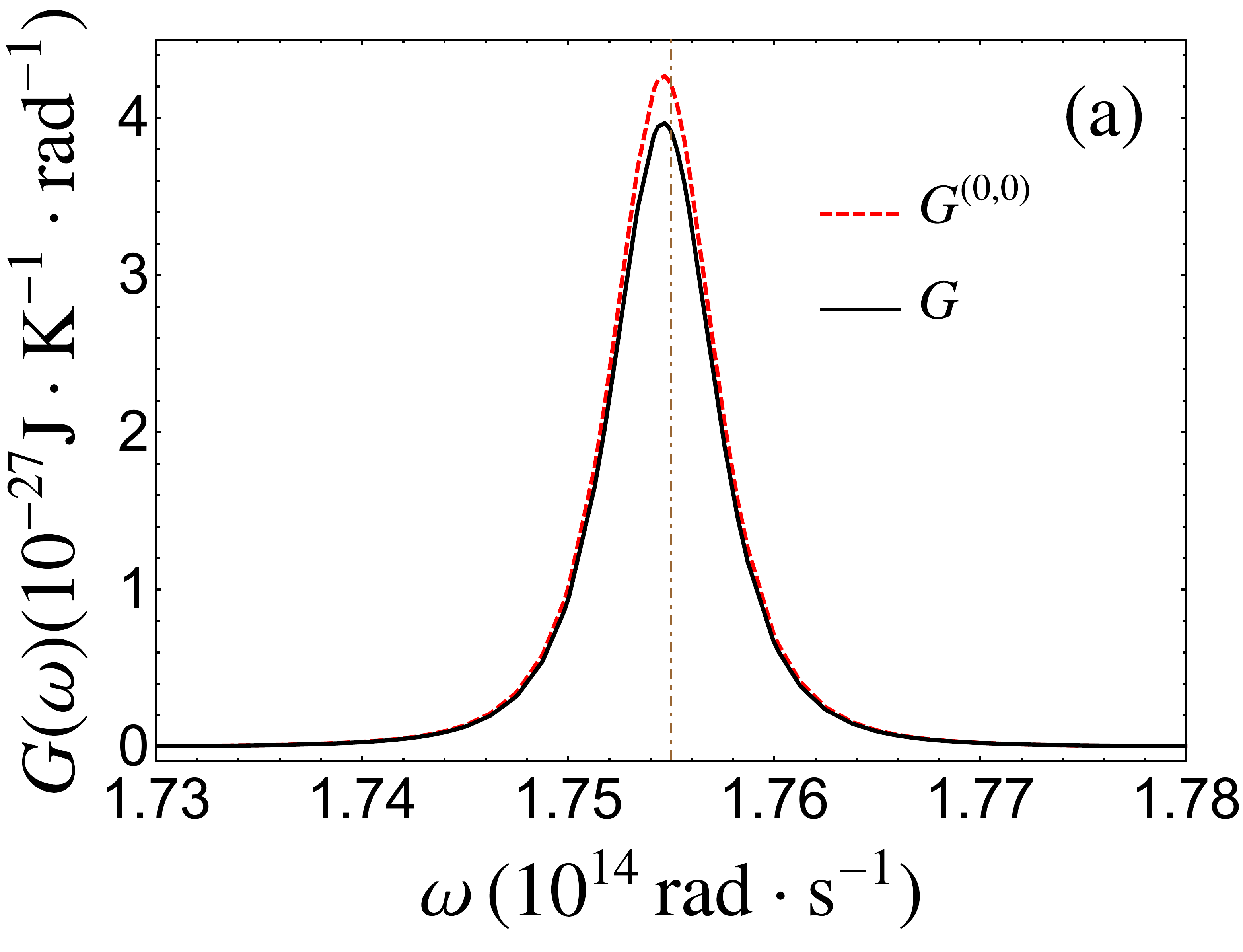}\quad\includegraphics[height=0.23\textwidth]{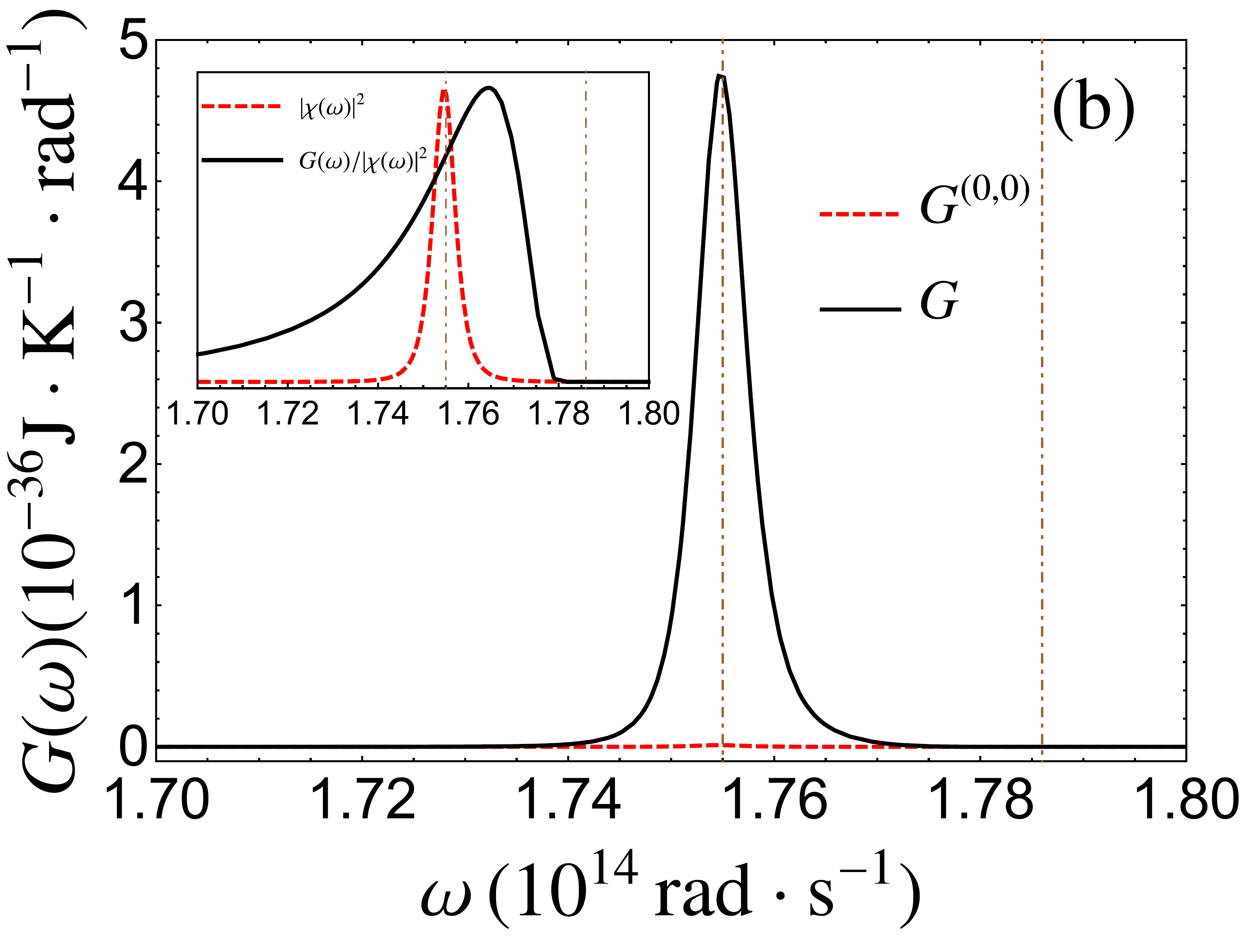}\quad\includegraphics[height=0.23\textwidth]{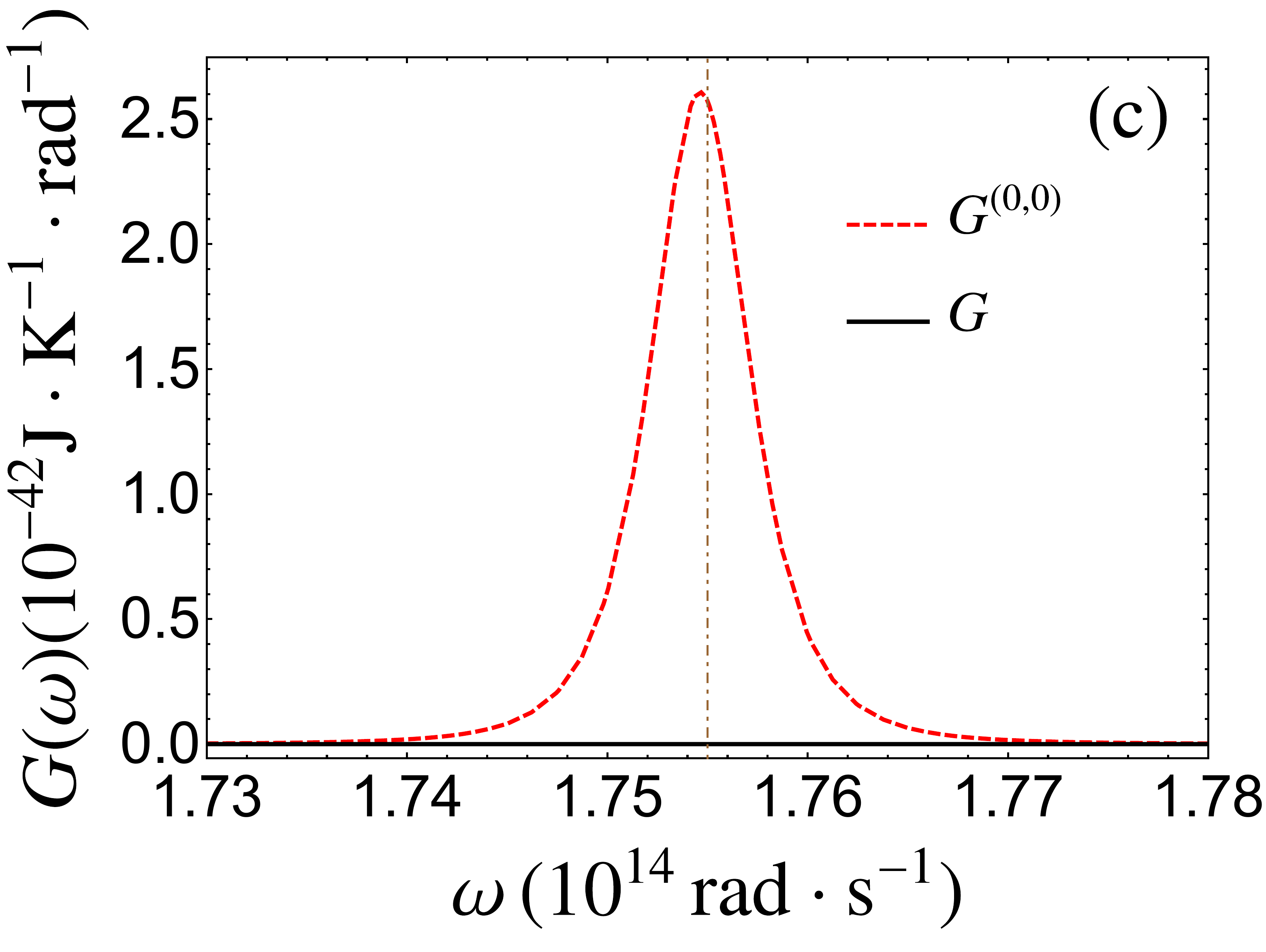}\\
\caption{Spectral conductance between two \textcolor{black}{SiC} nanoparticles at distance $d$, placed at distance \textcolor{black}{$z=50\,$nm} from a SiC substrate. Panels (a), (b) and (c) correspond to \textcolor{black}{$d=0.05,15,1000\,\mu$m} respectively. The dot-dashed vertical lines in each panel are associated with the nanoparticle and planar resonance frequencies, $\omega_\text{np} = 1.755\times 10^{14}\,$rad/s and $\omega_\text{pl} = 1.786\times 10^{14}\,$rad/s, respectively. The inset of panel (b) shows, in arbitrary units, $\chi^2$ (red dashed line) and $G/\chi^2$ (black solid line) as a function of frequency.}
\label{Spectrum_SiC}
\end{figure*}

We now dispose of all the tools needed to calculate the conductance in our system. As anticipated, we will mainly compare the conductance $G^{(0)}$ in vacuum to the total one $G$ in the presence of the interface. Concerning the latter, we choose a substrate made of silicon carbide (SiC), for two main reasons. First is that its dielectric properties can be easily described by using a Drude-Lorentz model~\cite{Palik}
\begin{equation}
\varepsilon(\omega)=\varepsilon_\infty \frac{\omega^2_L-\omega^2-i\Gamma\omega}{\omega^2_T-\omega^2-i\Gamma\omega},
\end{equation}
with \textcolor{black}{high-frequency} dielectric constant $\varepsilon_\infty=6.7$, longitudinal optical frequency $\omega_L=1.83\times 10^{14}\,$rad/s, transverse optical frequency $\omega_T=1.49\times 10^{14}\,$rad/s, and damping $\Gamma=8.97\times 10^{11}\,$rad/s. Moreover, according to this model the SiC--vacuum interface supports a \emph{planar} surface phonon-polariton mode in $p$ polarization at frequency $\omega_\text{pl} = 1.786\times 10^{14}\,$rad/s. This frequency corresponds to the resonance of the reflection coefficient $r_p$, condition which for large values of the wavevector $k$ is reduced to $\varepsilon(\omega) + 1 = 0$. It must be stressed here that the expression of the \textcolor{black}{electric} polarizability given in Eq.~\eqref{alphas} also predicts the existence of a surface resonance, which clearly differs from the one discussed above because of the different geometry of the interface. More specifically this second \emph{nanoparticle} resonance frequency $\omega_\text{np}$ corresponds \textcolor{black}{asymptotycally} to the condition $\varepsilon(\omega) + 2 = 0$, which for SiC gives $\omega_\text{np} = 1.755\times 10^{14}\,$rad/s. \textcolor{black}{It is well known that in the scenario of dielectric nanoparticles the electric contribution to the heat transfer (and thus to the conductance) dominates by orders of magnitude the magnetic one~\cite{POC2008}. For this reason, we will limit our discussion to the electric contribution in the case of SiC nanoparticles.}

We stress \textcolor{black}{that} in our configuration, compared to a standard calculation of two-body radiative heat transfer, the definitions of near and far field are more subtle. In fact, we have two distances, the particle-interface distance $z$ and the particle-particle distance $d$. When $d$ is small, we can expect the conductance to experience, even in vacuum, an amplification due to the nanoparticle surface mode. On the contrary, the distance $z$ is expected to regulate the participation of the substrate surface mode. This second distance is thus a more relevant parameter in our analysis of surface-wave-mediated modification of the heat transfer between the two nanoparticles.

\subsection{SiC nanoparticles on a SiC substrate}

We start our numerical analysis with the configuration in which both the nanoparticles and the substrate are made of SiC. In this case, we anticipate the participation (at least for some values of $z$ and $d$) of both the nanoparticle and the planar surface modes. Let us begin by discussing the conductance as a function of the interparticle distance $d$ for the \textcolor{black}{smallest} particle-surface distance \textcolor{black}{$z=50\,$nm}. In the plot shown in Fig.~\ref{SiC_01} we start by comparing the red dashed line, corresponding to the vacuum case $G^{(0)}$, to the black solid line, associated with the total value $G$. We clearly distinguish three zones with respect to the interparticle distance $d$. For very \textcolor{black}{small} distances the vacuum contribution dominates the scattered part and we have $G\simeq G^{(0,0)}$. In this \textcolor{black}{small}-distance region the particles are so close that their coupling is basically not influenced by the presence of the substrate. On the contrary, we observe a large region of $d$ in which not only is the role of the substrate relevant, but $G^{(\text{sc,sc})}$ becomes even the leading contribution to the total conductance $G$. More in detail, around $d=10\,\mu$m the conductance is dramatically amplified with respect to the vacuum configuration. As shown in the inset of Fig.~\ref{SiC_01}, showing the ratio between $G$ and $G^{(0,0)}$, the amplification goes beyond two orders of magnitude, reaching a value \textcolor{black}{close to 400}. Going back to the main part of Fig.~\ref{SiC_01}, we highlight a third region with respect to the distance $d$ ($d\gtrsim100\,\mu$m), for which the value of the conductance $G$ goes significantly below the vacuum result $G^{(0,0)}$. This means that for large values of the particle-particle distance the presence of the substrate inhibits the energy flux between the two nanoparticles. Although for these values of the distance the value of the conductance is very low, we can try to give a numerical description of the behavior of the different contributions to $G$. Starting with $G^{(0,0)}$, it can be easily shown from Eqs.~\eqref{ExprG} and \eqref{G0} that the low- and large-distance behaviors of the conductance are respectively $d^{-6}$ and $d^{-2}$. More specifically, we have
\begin{equation}\label{d-6-2}\begin{split}
 d&\to0,\quad G \simeq \frac{3\hbar}{4\pi^3 d^6}\int_0^{+\infty}\!\!d\omega\,\omega\,n'(\omega,T)\chi^2,\\
 d&\to+\infty,\quad G \simeq \frac{\hbar}{4\pi^3 c^4 d^2}\int_0^{+\infty}\!\!d\omega\,\omega^5\,n'(\omega,T)\chi^2.
\end{split}\end{equation}
These two asymptotic behaviors are shown explicitly in Fig.~\ref{SiC_01} by means of thin black solid lines. We clearly observe that for large values of $d$ the term $G^{\text{(sc,sc)}}$ behaves exactly as $G^{(0,0)}$ and, more interestingly, the crossed term $G^{(0,\text{sc})}$ shares (after an oscillatory behavior for intermediate distances) the same $d^{-2}$ behavior given in Eq.~\eqref{d-6-2}, with a prefactor $-2$ (its asymptotic behavior is shown in Fig.~\ref{SiC_01} as well) which exactly cancels the leading terms of the two remaining $G^{(0,0)}$ and $G^{(\text{sc},\text{sc})}$. As a consequence, we observe that the total conductance $G$ tends to zero faster than $G^{(0,0)}$ (we have shown numerically that it behaves as $d^{-4}$), and thus the ratio $G/G^{(0,0)}$ tends to zero as $d^{-2}$. However, we remark again that this behavior takes place at values of the distance for which both $G^{(0,0)}$ and $G$ are very low.

\begin{figure}[t!]
\includegraphics[width=0.5\textwidth]{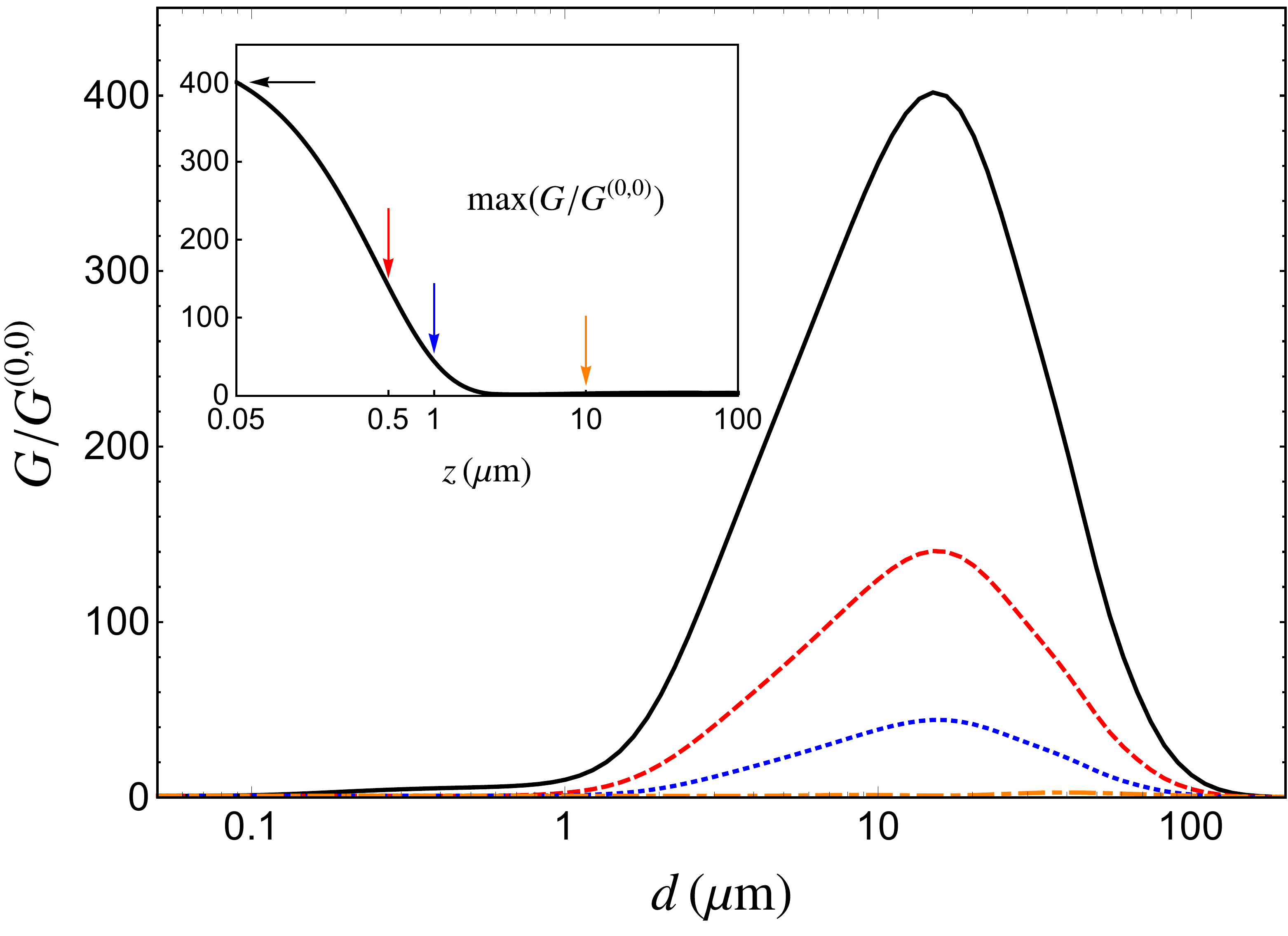}
\caption{The inset shows, as a function of the nanoparticles--surface distance $z$, the maximum of the ratio $G/G^{(0,0)}$ with respect to the interparticle distance $d$ for two SiC nanoparticles on top of a SiC substrate. In the main part of the figure, the same ratio is plotted as a function of $d$ for \textcolor{black}{$z=50\,$nm} (black solid line), 500\,nm (red dashed line), $1\,\mu$m (blue dotted line) and $10\,\mu$m (orange dot-dashed line).}
\label{SiC_z}
\end{figure}

In order to get more insight into this amplification mechanism, we turn our attention to a spectral analysis of the conductance, by looking at the function $G(\omega)$ such that $G=\int_0^{+\infty} d\omega\,G(\omega)$. In particular we focus on the configuration \textcolor{black}{$z=50\,$nm} discussed so far and shown in Fig.~\ref{SiC_01} and plot $G(\omega)$ for the three values of \textcolor{black}{$d=0.05,15,1000\,\mu$m} corresponding to the extreme values taken into account in Fig.~\ref{SiC_01}, along with the intermediate value ($d=15\,\mu$m), close to the one corresponding to the maximum amplification $G/G^{(0,0)}$. In Fig.~\ref{Spectrum_SiC}(a) we clearly see that for \textcolor{black}{$d=50\,$nm} the vacuum contribution $G^{(0,0)}$ gives almost the entire value of the conductance $G$. Moreover, we see that the spectral conductance $G(\omega)$ is clearly resonating at the nanoparticle resonance frequency $\omega_\text{np}$ (the vertical dot-dashed line in the figure), coherently with the \textcolor{black}{small} distance between the two particles. For the other extreme value of the distance $d=1000\,\mu$m, Fig.~\ref{Spectrum_SiC}(c) shows that the vacuum contribution dominates with respect to $G$, coherently with the different power law discussed above. In terms of spectral contributions, $G^{(0,0)}$ is again peaked at $\omega_\text{np}$, which is the only resonance frequency defined in the absence of the interface. We are left with the more subtle discussion of the intermediate regime, corresponding to the case $d=15\,\mu$m shown in Fig.~\ref{Spectrum_SiC}(b). In the main part of the figure, we clearly see that $G$ is much larger than $G^{(0,0)}$, showing that we are indeed observing a conductance amplification induced by the presence of the substrate. Nevertheless, the total spectral conductance $G(\omega)$ is still peaked at $\omega_\text{np}$, as in the \textcolor{black}{small}-distance configuration \textcolor{black}{$d=50\,$nm} shown in panel (a). This is somehow surprising since, based on the interpretation of this amplification as an effect of the surface wave existing at the planar vacuum-SiC interface, we would have expected a resonance at the larger frequency $\omega_\text{pl}$ [the second vertical line shown in Fig.~\ref{Spectrum_SiC}(b)]. In order to delve deeper into this behavior, we remark that the spectral conductance $G(\omega)$ is the product of the term $\chi^2$ [see Eq.~\eqref{ExprG}], the only one depending on the two nanoparticles and resonating at $\omega_\text{np}$, and of an expression involving $n'(\omega,T)$ and $\Tr\bigl(\mathds{G}\mathds{G}^\dag\bigr)$. These two contributions are represented, in arbitrary units, in the inset of Fig.~\ref{Spectrum_SiC}(b). We see that, while $\chi^2$ is clearly \textcolor{black}{peaked} at $\omega_\text{np}$, the remaining multiplicative factor has a broader peak at a frequency larger than $\omega_\text{np}$ but still smaller than the expected one $\omega_\text{pl}$. This is a result of the complicated interplay between the resonance of the reflection coefficient $r_p$ and the oscillatory behavior of the Bessel functions appearing in Eq.~\eqref{Gsc} as a function of $k$. These oscillations partially cancel the resonant behavior of $r_p$ for large values of $k$ (for which the resonance frequency approaches asymptotically $\omega_\text{pl}$) and produce the broader peak shown in the inset of Fig.~\ref{Spectrum_SiC}(b). Finally, the product of this term and $\chi^2$ is at the origin of the maximum of $G(\omega)$ at $\omega_\text{np}$ even in the case of maximum amplification.

We finally address the question of the dependence of the conductance amplification on the distance $z$ between the two nanoparticles and the substrate. In the inset of Fig.~\ref{SiC_z} we plot as a function of $z$ (up to $100\,\mu$m) the maximum of the ratio $G/G^{(0,0)}$ with respect to $d$ in the range \textcolor{black}{$[50\,\text{nm},1000\,\mu\text{m}]$}. We clearly recognize the value around \textcolor{black}{400} for \textcolor{black}{$z=50\,$nm} and we observe a \textcolor{black}{monotonically} decreasing behavior as a function of $z$. Remarkably, the amplification factor is still 10 around \textcolor{black}{$z=1.6\,\mu$m}. Moreover, the decay rate of the maximum as a function of $z$ is of the order of some microns, comparable to the decay length of the phonon-polariton at the SiC--vacuum interface. We finally show in the main part of Fig.~\ref{SiC_z}, for some values of $z$ (\textcolor{black}{$z=0.05,0.5,1,10\,\mu$m}) the ratio $G/G^{(0,0)}$ as a function of the interparticle distance $d$. Apart from the expected decay in the peak height when increasing the value of $z$, we observe that up to $z\simeq1\,\mu$m the distance $d$ realizing the maximum amplification is approximately unchanged around $d\simeq15\,\mu$m, while for $z=10\,\mu$m it moves to higher values of $d$, where the vacuum conductance $G^{(0,0)}$ is already much smaller, making the amplification mechanism (already less pronounced) even less interesting.

\subsection{Gold nanoparticles on a SiC substrate}

\begin{figure}[t!]
\includegraphics[width=0.5\textwidth]{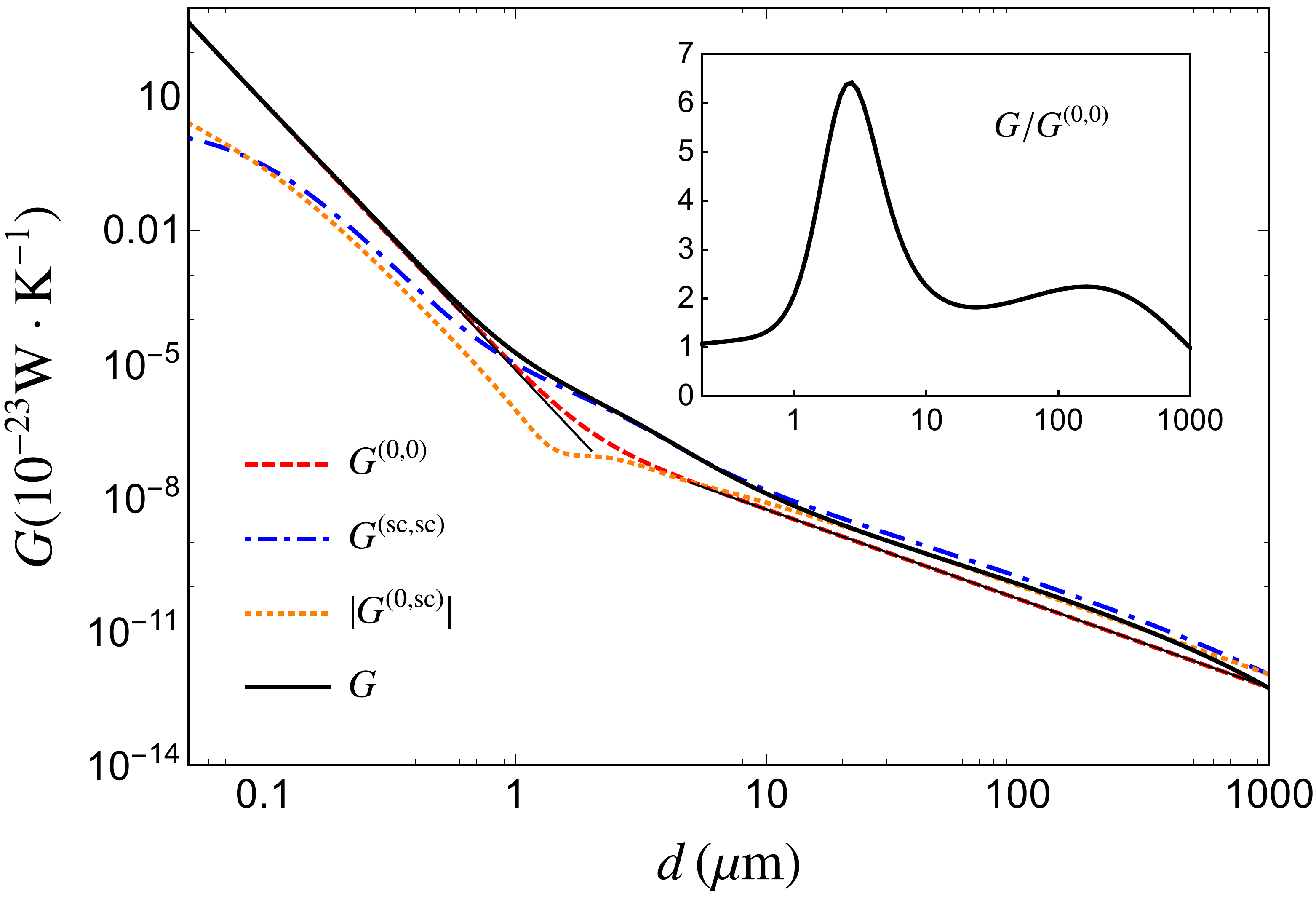}
\caption{Total conductance $G$ (black solid line) and single contributions $G^{(0,0)}$ (red dashed line), $G^{(\text{sc,sc})}$ (blue dot-dashed line) and $G^{(\text{0,sc})}$ (in absolute value, orange dotted line) as defined in Eq.~\eqref{ContributionsG} between two gold nanoparticles at distance $d$, placed at distance \textcolor{black}{$z=50\,$nm} from a SiC substrate. The thin black lines correspond to the \textcolor{black}{small}- and long-distance asymptotic behaviors of $G^{(0,0)}$ given in Eq.~\eqref{d-6-2}. The inset shows the ratio between conductances in the presence and absence of substrate as a function of $d$.}
\label{Au_01}
\end{figure}

\begin{figure*}[t!]
\includegraphics[height=0.23\textwidth]{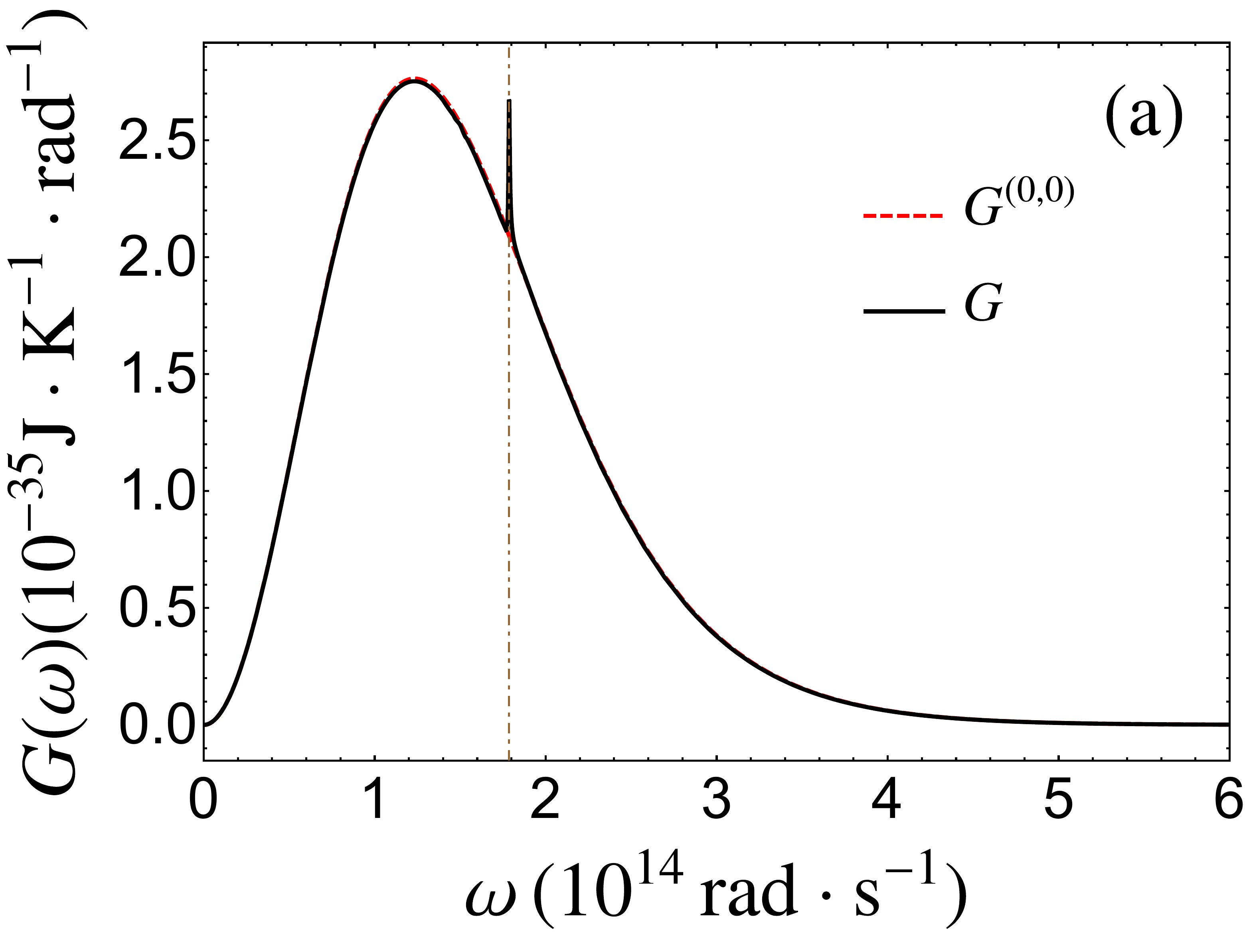}\quad\includegraphics[height=0.23\textwidth]{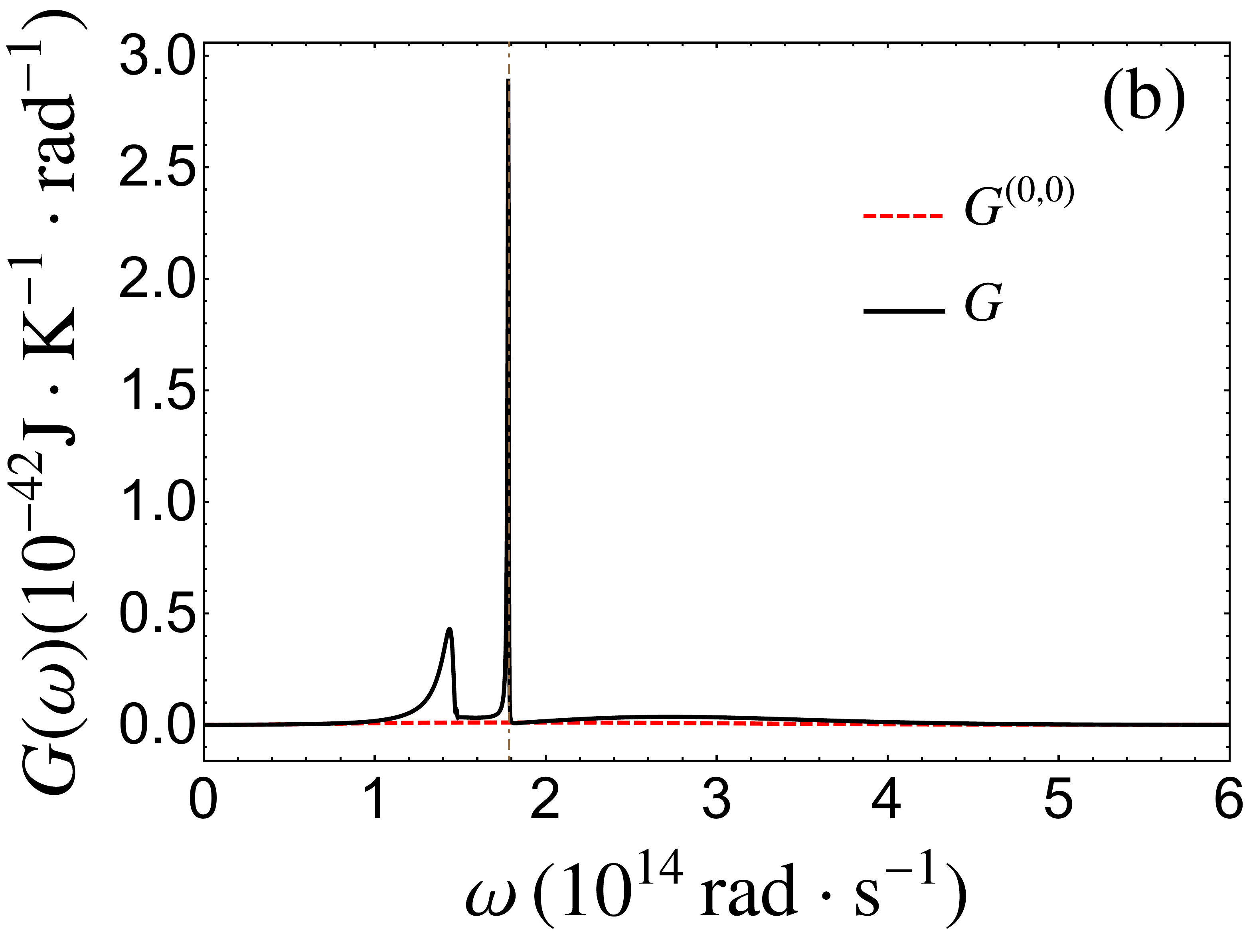}\quad\includegraphics[height=0.23\textwidth]{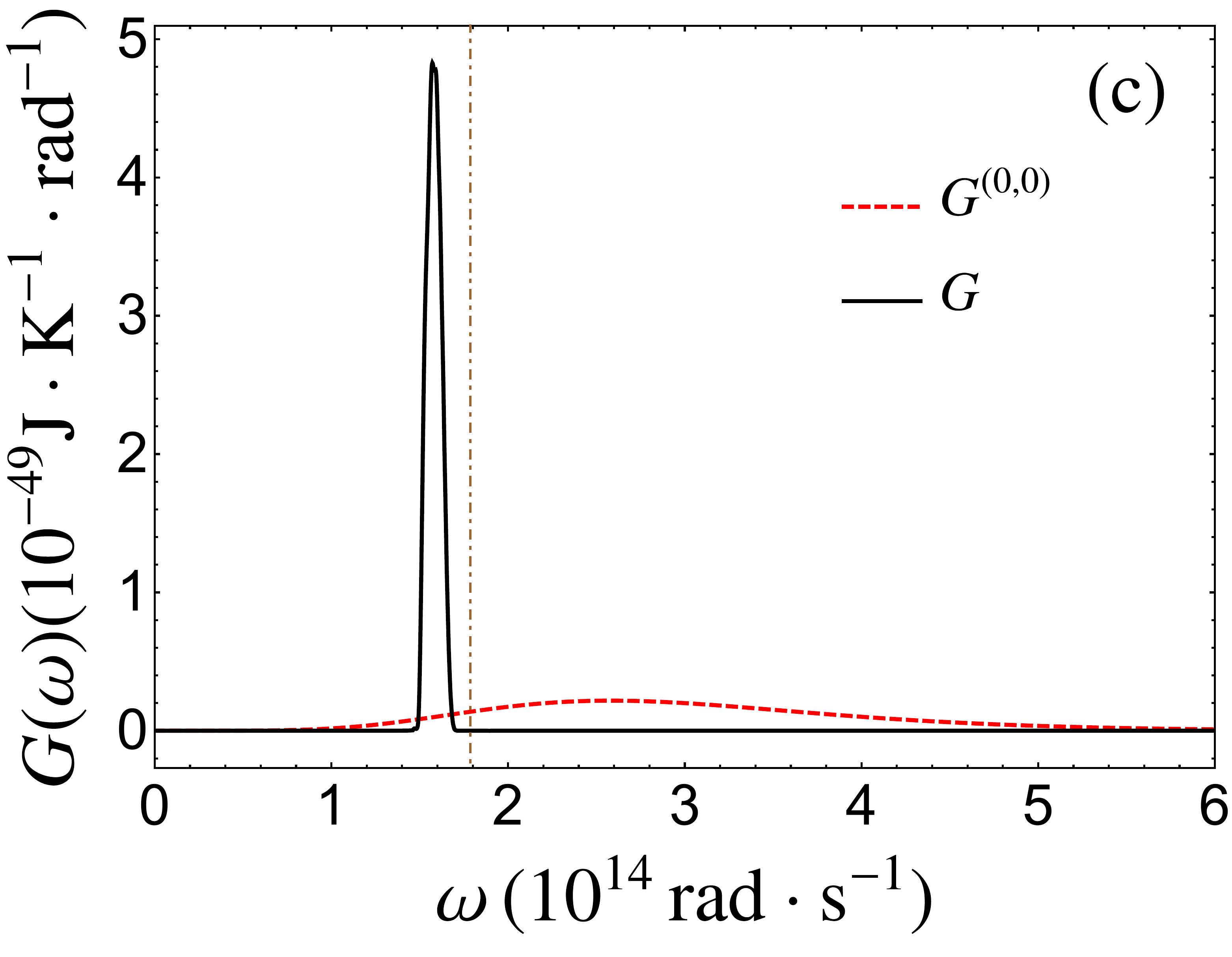}\\
\caption{Spectral conductance between two \textcolor{black}{gold} nanoparticles at distance $d$, placed at distance \textcolor{black}{$z=50\,$nm} from a SiC substrate. Panels (a), (b) and (c) correspond to \textcolor{black}{$d=0.05,2,1000\,\mu$m} respectively. The dot-dashed vertical line in each panel is associated with the planar resonance frequency $\omega_\text{pl} = 1.786\times 10^{14}\,$rad/s, respectively.}
\label{Spectrum_Au}
\end{figure*}

We now turn our attention to a different configuration, in which the substrate is again made of SiC, thus supporting the same surface wave discussed so far, while the two particles are made of gold. \textcolor{black}{To describe the permittivity of gold we use a modified Drude model~\cite{Gold}
\begin{equation}
\varepsilon(\omega)= 1 - \frac{\omega^2_P}{\omega[\omega + i\Gamma(1+v_F/R)]},
\end{equation}
with plasma frequency $\omega_P=1.71\times 10^{16}\,$rad/s and dissipation rate $\Gamma=4.05\times 10^{13}\,$rad/s. The term proportional to the Fermi velocity $v_F=1.2\times10^6\,\text{m}/\text{s}$ takes into account the deviation from the bulk permittivity associated with the small size of the particle~\cite{POC2008,Gold}.} In this case, a nanoparticle surface mode [a zero of $\varepsilon(\omega)+2$] still exists, but its frequency \textcolor{black}{(in the ultraviolet range)} is such that it does not take part to the energy exchange since it falls beyond the frequency window fixed by the function $n'(\omega,T)$ in Eq.~\eqref{ExprG}. \textcolor{black}{In the case of gold nanoparticles, it was shown that the magnetic contribution to the interparticle energy flux typically dominates over the electric one~\cite{POC2008}. More specifically, for our choice of radius $R=5\,$nm, this is still true but the two contributions can be comparable. For this reason, in what follows we include both terms in our calculation, but we neglect for simplicity crossed electric-magnetic terms~\cite{Add8}.}

Based on this model we start computing the total conductance $G$ (as well as all the individual contributions) as a function of $d$ for \textcolor{black}{$z=50\,$nm}. The results are shown in Fig.~\ref{Au_01}, and the comparison with Fig.~\ref{SiC_01} shows a dramatic reduction of the values of $G$, because of the replacement of dielectric with metallic nanoparticles. Nevertheless, even in this configuration an amplification of the conductance with respect to the vacuum configuration is possible. In particular, as shown in the inset, we obtain a maximum \textcolor{black}{larger than 6} for a distance $d\simeq2\,\mu$m. For larger distances, while the asymptotic $d^{-2}$ is still visible, the interplay with the terms $G^{(\text{sc,sc})}$ and $G^{(0,\text{sc})}$ is less manifest in the interval of distances under scrutiny.

We repeat here the spectral analysis of the conductance for \textcolor{black}{$d=0.05,2,1000\,\mu$m}, i.e. respectively the lower boundary, the location associated with the maximum amplification, and the upper boundary. The results are shown in Fig.~\ref{Spectrum_Au}. In panel (a) we observe that for \textcolor{black}{$d=50\,$nm} the vacuum result shows a flat spectrum, coherently with the absence of a nanoparticle resonance in the frequency window relevant at $T=300\,$K. On the contrary, including the substrate results in the appearance of a narrow peak at the planar SiC--vacuum resonance frequency $\omega_\text{pl}$, but this peak produces here a negligible amplification. In Fig.~\ref{Spectrum_Au}(b), we show the case $d=2\,\mu$m, where the amplification is maximized. A similar peak appears in the presence of the substrate, \textcolor{black}{along with a broader peak at lower frequencies coming from the magnetic contribution. These are }responsible of the predicted amplification. Finally, for large $d$ we still find a signature of the SiC--vacuum surface mode, but with an associated value of the amplification again around 1, as in the case \textcolor{black}{$d=50\,$nm}.

\begin{figure}[b!]
\includegraphics[width=0.5\textwidth]{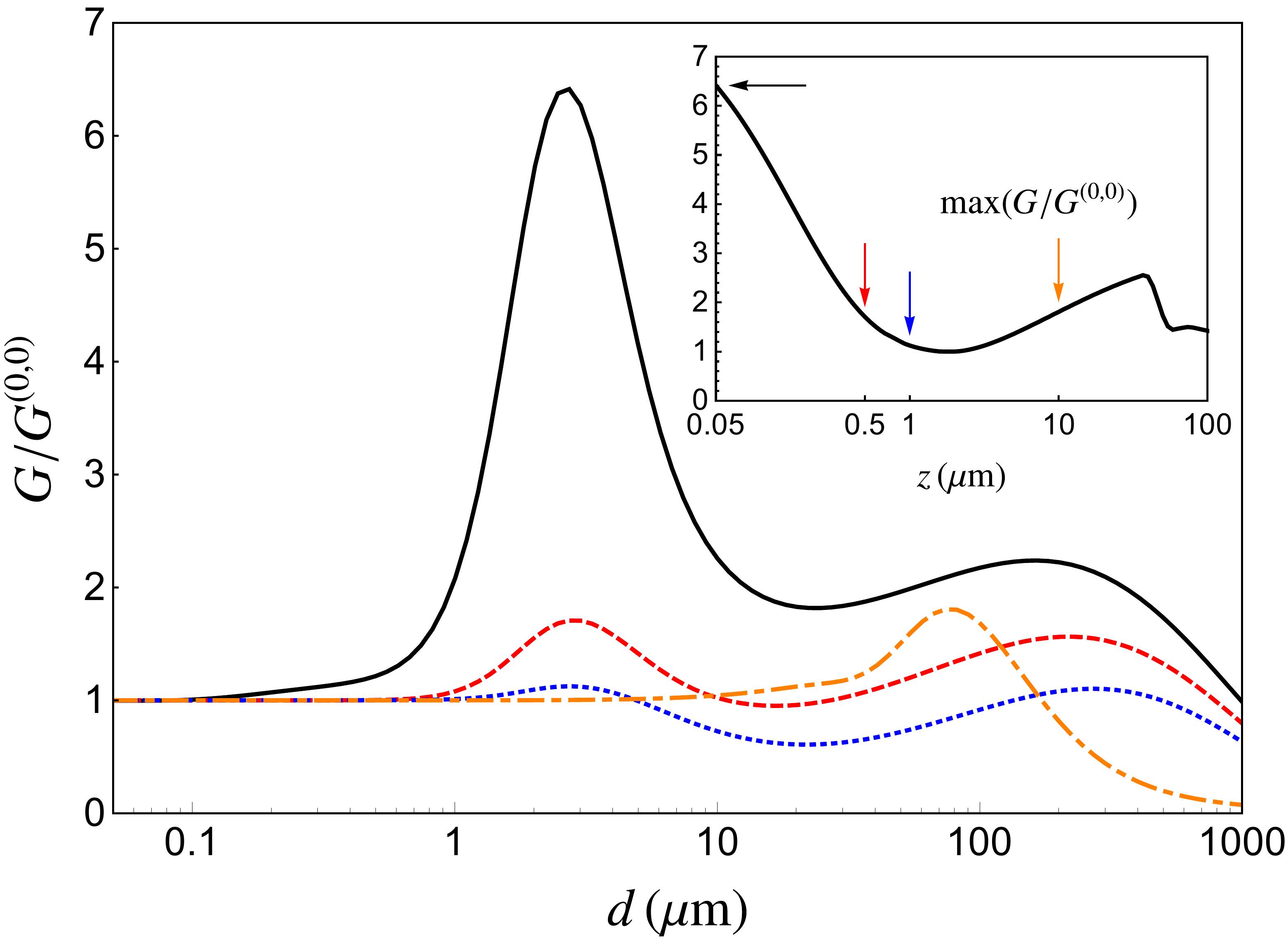}
\caption{The inset shows, as a function of the nanoparticles--surface distance $z$, the maximum of the ratio $G/G^{(0,0)}$ with respect to the interparticle distance $d$ for two gold nanoparticles on top of a SiC substrate. In the main part of the figure, the same ratio is plotted as a function of $d$ for \textcolor{black}{$z=50\,$nm} (black solid line), 500\,nm (red dashed line), $1\,\mu$m (blue dotted line) and $10\,\mu$m (orange dot-dashed line).}
\label{Au_z}
\end{figure}

We discuss now the dependence of the amplification mechanism on $z$. The inset of Fig.~\ref{Au_z} shows a more complicated dependence than in the case of SiC (see Fig.~\ref{SiC_z}), since the initial similar decay of the maximum possible ratio $G/G^{(0,0)}$ is followed by an increase from approximately 1 to $20\,\mu$m. Nevertheless, as shown in the main part of Fig.~\ref{Au_z}, while during the initial decrease of $G/G^{(0,0)}$ as a function of $z$, the value of $d$ realizing the maximum remains approximately constant around $d\simeq2\,\mu$m, it then increases during the ascending part, moving to values at which the overall conductance becomes tiny and thus the amplification less \textcolor{black}{relevant}.

\section{Role of a graphene sheet}\label{SecGraphene}

We have shown so far that the presence of a surface wave at the interface between vacuum and the substrate is able to produce an amplification of the conductance between the two nanoparticles. This amplification goes beyond two orders of magnitude when both the particles and the substrate are made of SiC, while it is \textcolor{black}{around 6} in the case of gold nanoparticles on top of a SiC substrate. The reason of this dramatic difference is that not only SiC nanoparticles do support a surface mode in the Planck  window which participates to the energy exchange, but the frequency of the surface mode existing at the interface is almost resonant with the first one. This is not the case for gold nanoparticles, which support a surface mode as well, but in the ultraviolet range of frequencies. For this reason, it would be interesting to tailor the interplay between the planar surface mode and the nanoparticles in order to reduce the mismatch between the two resonance frequencies. A remarkable recent interest has been focused on the use of graphene to manipulate (both spectrally and in terms of absolute value) near-field radiative heat transfer~\cite{Persson,Volokitin,Svetovoy1,Ilic1,Ilic2,Messina1,Messina2,Lim2,Phan,Liu3,Liu1,Svetovoy2,Drosdoff,Zhang1,Lim1,Chang,Song,Yin,Zheng,Zhao1,Simchi,Lim3,Zhao2,Shi,BenAbdallahAPL15,Messina3,Zhao}. By considering both monolayer and multilayer structures, it has been shown that the remarkable optical properties of graphene~\cite{Geim1,Geim2} can be exploited in order to actively control the radiative heat transfer, also thanks to the possible manipulation of the optical response of graphene by means of a modification of its chemical potential. Recently it has been shown that the F\"{o}rster resonance energy transfer between two particles can be enhanced by up to six orders of magnitude~\cite{BiehsAPL13}.

In this Section, we explore how the presence of a graphene sheet deposited on the SiC substrate influences the amplification effect observed so far. \textcolor{black}{The optical properties of graphene are conveniently described in terms of a conductivity $\sigma$. For this quantity we will employ a description based on the sum of an intraband (Drude) and an interband contribution, given by \cite{FalkovskyJPhysConfSer08}
\begin{eqnarray}\label{Sigma}&&\sigma_D(\omega)=\frac{i}{\omega+\frac{i}{\tau}}\frac{2e^2k_BT}{\pi\hbar^2}\log\Bigl(2\cosh\frac{\mu}{2k_BT}\Bigr),\\
&&\nonumber\sigma_I(\omega)=\frac{e^2}{4\hbar}\Bigl[G\Bigl(\frac{\hbar\omega}{2}\Bigr)+i\frac{4\hbar\omega}{\pi}\int_0^{+\infty}\frac{G(\xi)-G\bigl(\frac{\hbar\omega}{2}\bigr)}{(\hbar\omega)^2-4\xi^2}\,d\xi\Bigr],\end{eqnarray}
where $G(x)=\sinh(x/k_BT)/[\cosh(\mu/k_BT)+\cosh(x/k_BT)]$. The conductivity depends explicitly on the temperature $T$ of the graphene sheet, for which we have chosen $T=300\,$K in our calculations. Moreover, it contains the chemical potential $\mu$, which represents an adjustable parameter allowing us to actively tune the optical properties. Finally, for the relaxation time $\tau$ we have chosen the value~\cite{JablanPRB09} $\tau=10^{-13}\,$s. The knowledge of the conductiviy allows us to write the expression of the Fresnel reflection coefficients, modified by the presence of the graphene sheet. These new coefficients read~\cite{Messina2}
\begin{equation}\begin{split}
r_s &=\frac{k_z-k_{zm}-\mu_0\sigma(\omega)\omega}{k_z+k_{zm}+\mu_0\sigma(\omega)\omega},\\
r_p &= \frac{\varepsilon_0\omega[\varepsilon(\omega)k_z-k_{zm}]+\sigma(\omega)k_zk_{zm}}{\varepsilon_0\omega[\varepsilon(\omega)k_z+k_{zm}]+\sigma(\omega)k_zk_{zm}},
\end{split}\end{equation}
and need to be used in the calculation of the Green's function given by Eq.~\eqref{Gsc}.}

\begin{figure}[t!]
\includegraphics[width=0.495\textwidth]{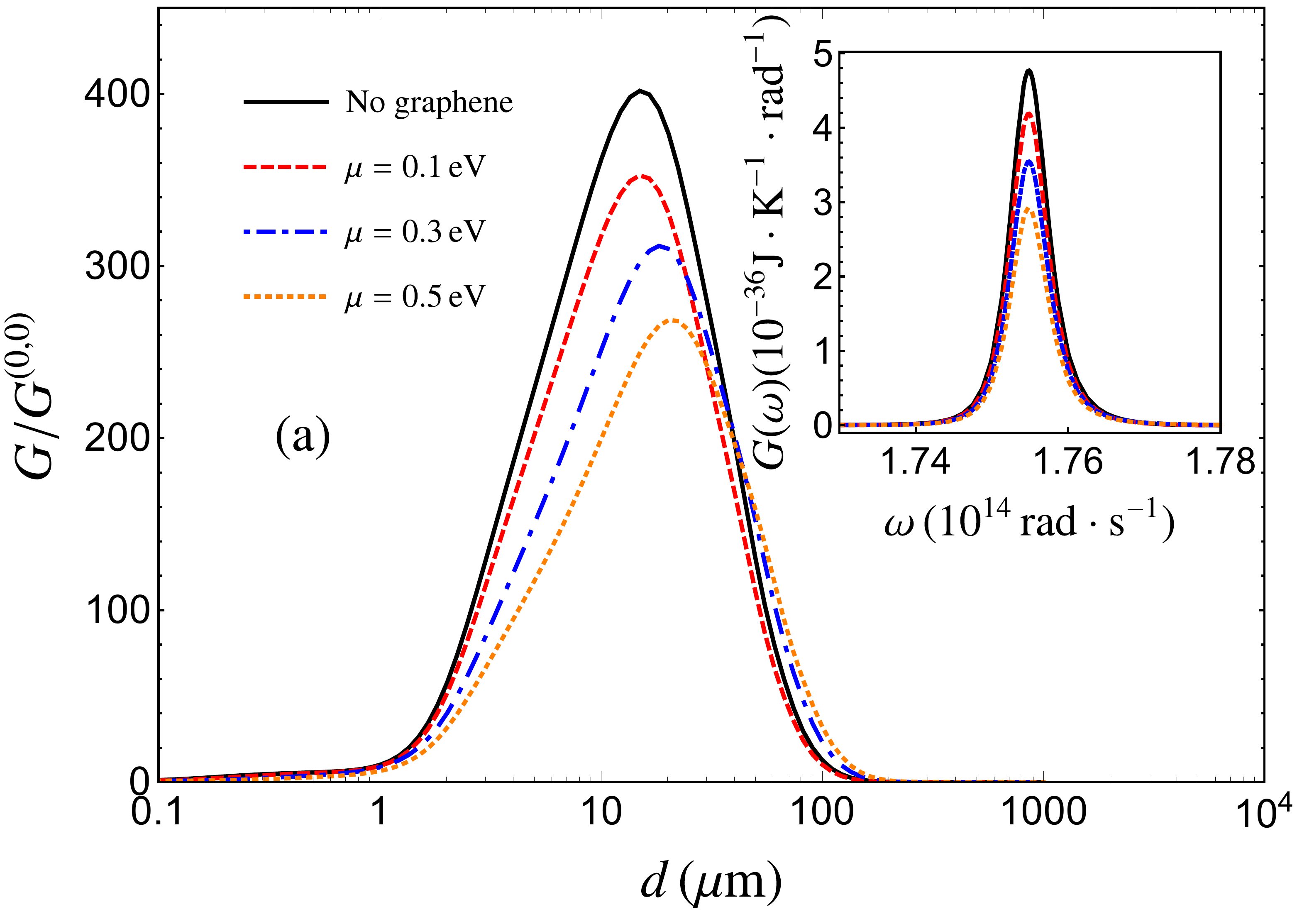}
\includegraphics[width=0.5\textwidth]{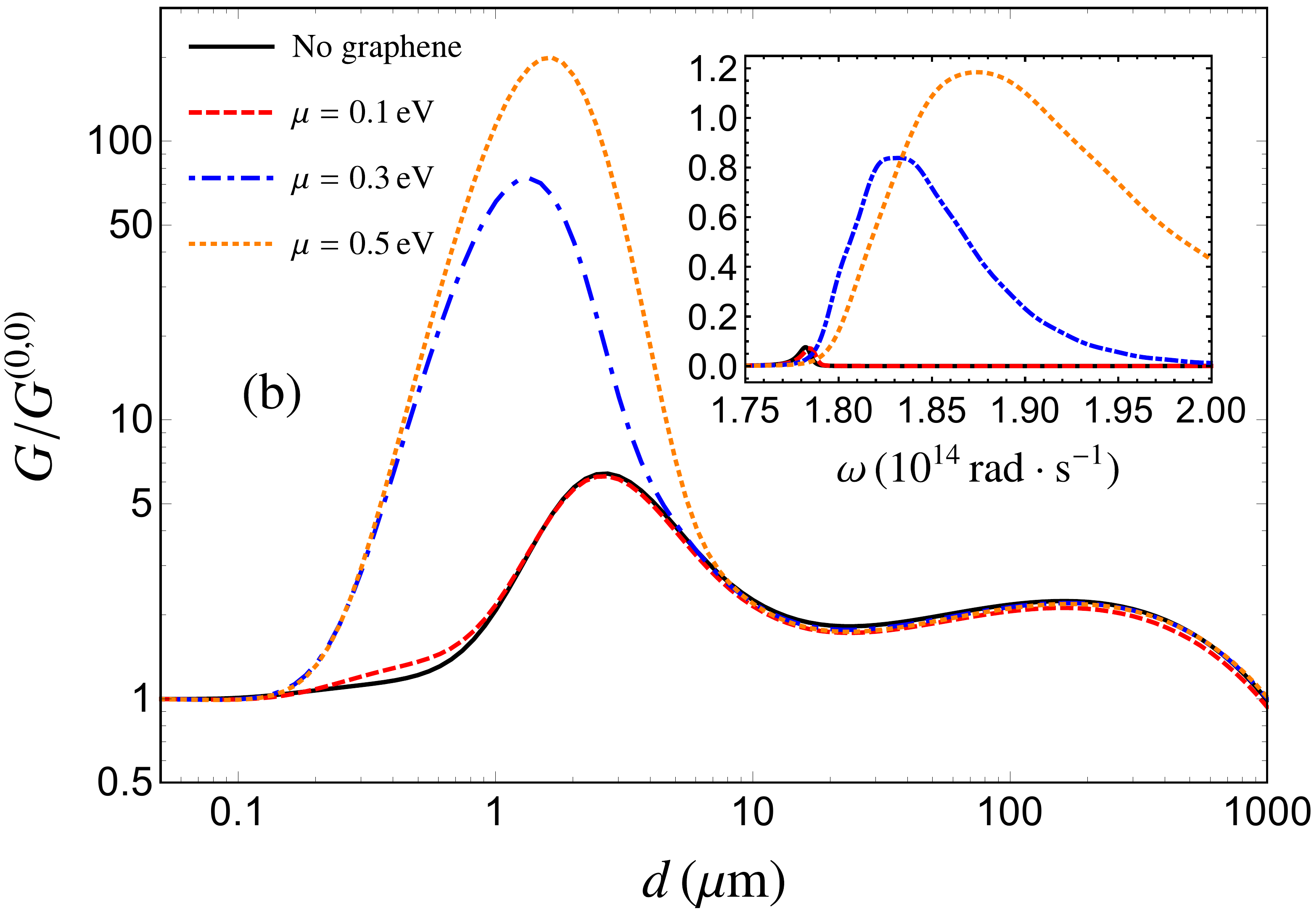}
\caption{Panel (a) shows the conductance ratio $G/G^{(0,0)}$ as a function of $d$ between two SiC nanoparticles placed at distance \textcolor{black}{$z=50\,$nm} from a SiC substrate. The four lines correspond to the absence of graphene (black solid line), and to configurations with graphene having $\mu=0.1\,$eV (red dashed line), 0.3\,eV (blue dot-dashed line) and 0.5\,eV (orange dotted line). The inset shows the spectral conductance associated with the four same configurations. Panel (b) and its inset show the same quantities for two gold nanoparticles placed at distance \textcolor{black}{$z=50\,$nm} from a SiC substrate.}
\label{Graphene}
\end{figure}

\textcolor{black}{In order to get an insight on the possibilities offered by the presence of a graphene sheet,} we focus on the best scenario with respect to the particle-surface distance, namely \textcolor{black}{$z=50\,$nm}, and plot as a function of $d$ the conductance ratio $G/G^{(0,0)}$ both for SiC and gold nanoparticles in Fig.~\ref{Graphene}. For both configurations, the case in the absence of graphene is compared with scenarios in the presence of graphene having the three different chemical potentials $\mu=0.1,0.3,0.5\,$eV. Let us begin by analyzing the results for SiC particles, shown in Fig.~\ref{Graphene}(a). In this case, we observe that the presence of graphene reduces the overall conductance amplification. Moreover, this reduction increases when increasing the graphene chemical potential. This is coherent with our previous description of the local density of states in the vicinity of a graphene-covered substrate, described in Ref.~\onlinecite{Messina1}. The surface plasmon supported by graphene alone couples with the phonon-polariton existing at the SiC--vacuum interface, producing a hybrid mode: this mode has a modified dispersion relation which no longer shows a horizontal frequency asymptote in the $(\omega,k)$ plane, but is shifted for any value of $k$ toward higher frequencies compared to the SiC--vacuum phonon-polariton alone. This modification reduces the coupling with SiC nanoparticles, thus reducing the conductance amplification. The inset of Fig.~\ref{Graphene}(a) shows a spectral analysis of this phenomenon at $d=15\,\mu$m. The existence of a peak at the nanoparticle surface resonance frequency is still a signature of the presence of the two nanoparticles, while the reduction of the height of the peak indicates the reduced coupling due to the presence of graphene.

It is interesting to move now to the case of gold nanoparticles, in which for the same reasons we expect the presence of graphene to increase the coupling with the SiC substrate. This is indeed the case, as we show in Fig.~\ref{Graphene}(b). While the lowest chemical potential $\mu=0.1\,$eV taken into account produces a result basically indistinguishable with respect to the absence of graphene, the two higher values of $\mu$ produce indeed an amplification of the effect, with a ratio $G/G^{(0,0)}$ going up to two orders of magnitude for $\mu=0.5\,$eV. We limit our analysis here to this value of $\mu$ for practical reasons, since a higher chemical potential could be challenging to produce experimentally, but our results show clearly that graphene allows in this scenario as well to widely tailor the amplification of conductance. In order to get a final insight into this effect, we plot in the inset of Fig.~\ref{Graphene}(b) the spectral conductance for $d=1.5\,\mu$m, corresponding to the maximum in Fig.~\ref{Graphene}(b) for $\mu=0.5\,$eV. While in the absence of graphene and for $\mu=0.1\,$eV we still see a signature of the planar surface mode at the SiC--vacuum interface, this is completely lost for higher values of the chemical potential, for which $G(\omega)$ shows the same broadening (associated with the modified dispersion relation) observed in Ref.~\cite{Messina1}.

\begin{figure}[t!]
\includegraphics[width=0.5\textwidth]{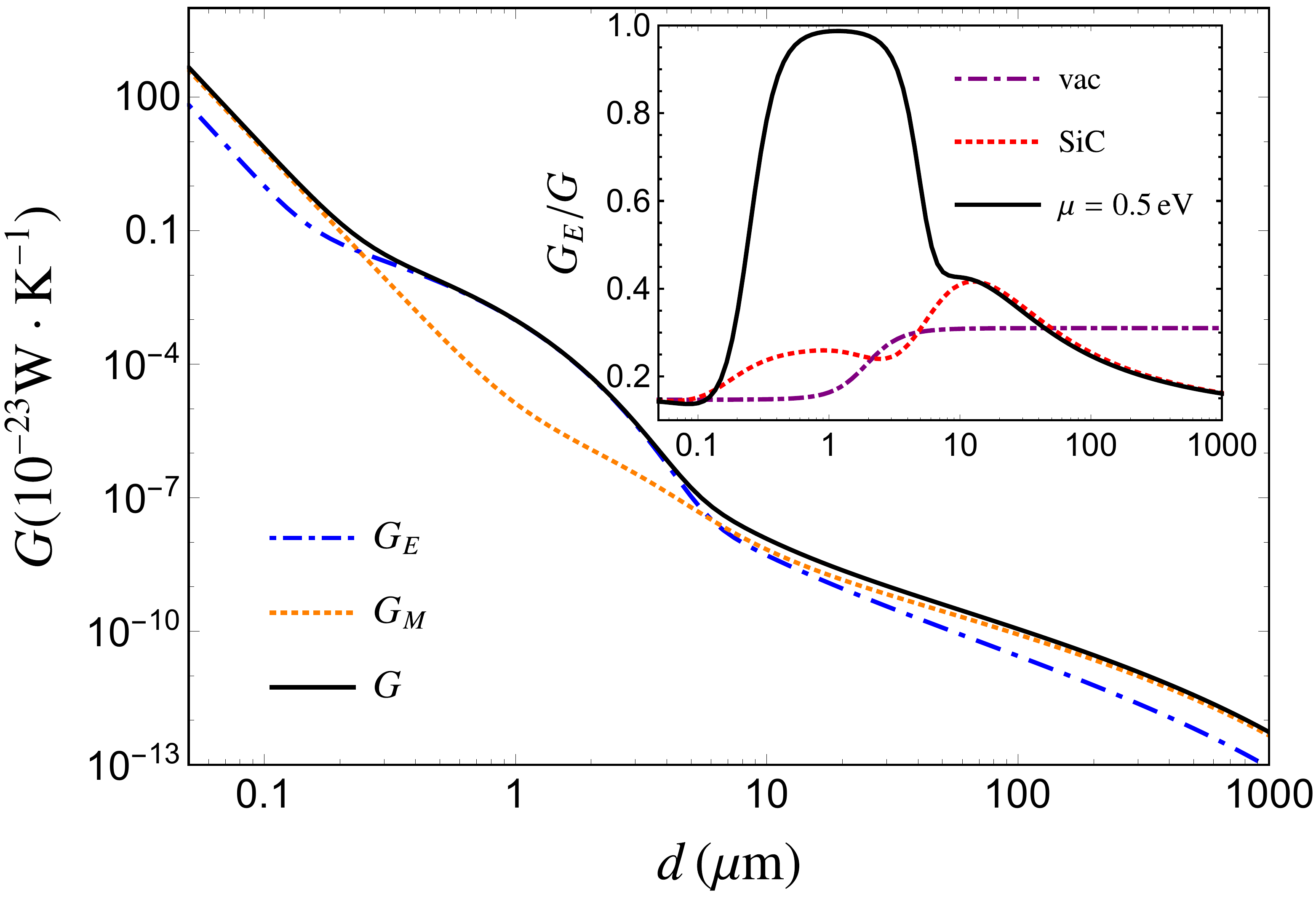}
\caption{\textcolor{black}{The main part of the figure shows the conductance (black solid line) between two gold particles placed at distance $z=50\,$nm from a SiC substrate covered with a graphene sheet having $\mu=0.5\,$eV. This is compared with the electric (blue dot-dashed line) and magnetic (orange dotted line) contributions. The inset shows the ratio between the electric contribution to the conductance and the total conductance in the same configuration (black solid line), for two gold particles in vacuum (purple dot-dashed line) and on top of a SiC substrate (red dotted line).}}
\label{GGE}
\end{figure}

\textcolor{black}{It is interesting to address one last point in the case of metallic nanoparticles, namely the interplay between electric and magnetic contribution in the presence of a graphene sheet. As a matter of fact, we can anticipate from the results of Ref.~\onlinecite{Messina1} that the presence of graphene modifies the electric part of the interaction. In order to verify this intuition, we plot in the main part of Fig.~\ref{GGE} the total conductance $G$ (black solid line) in the case of a graphene sheet with $\mu=0.5\,$eV, together with the two electric and magnetic contributions $G_E$ (blue dot-dashed line) and $G_H$ (orange dotted line). We clearly see that, while for small and large $d$ we observe that the magnetic contribution is slightly larger than the electric one (as described in Ref.~\onlinecite{POC2008}), the amplification associated with the presence of graphene entirely acts on the electric part, which dominates over the magnetic one for intermediate distnaces close to $1\,\mu$m. In the inset of Fig.~\ref{GGE} we plot the ratio $G_E/G$ in the same configuration (black solid line), confirming that $G_E$ basically coincides with $G$ in this intermediate-distance region. On the contrary, when the two gold particles are in vacuum (purple dot-dashed line) or on top of a SiC substrate (red dotted line), the electric part gives a contribution which varies between 10\% and 40\% of the total one.}

\section{Conclusions}\label{SecConclusions}

We have studied the modifcation of radiative heat exchanges between two dielectric (SiC) or metallic (gold) nanoparticles when placed in proximity of a dielectric (SiC) substrate supporting a surface phonon-polariton. We have shown that in both scenarios the presence of a surface wave can indeed amplify the ambient-temperature conductance between the nanoparticles. This happens by a factor of more than two orders of magnitude in the case of SiC particles, whereas the amplification factor is limited to \textcolor{black}{6} for gold nanoparticles. We have spectrally analyzed the effect, clearly highlighting the role played by the surface mode. Moreover, we have studied the dependence of the effect on the nanoparticle--surface distance $z$, showing that the effect is lost for large distances, as expected since surface modes are confined in the vicinity of the interface. Furthermore, we have shown that the presence of a graphene sheet on top of the substrate can dramatically modify and allow to tailor the amplification. In particular, in the case of gold nanoparticles, the reduction of frequency mismatch between the substrate and nanoparticle resonances allows to obtain in this configuration as well an amplification of two orders of magnitude.

Our work represents a first step in the study of modification of energy exchanges mediated by an interface and it certainly paves the way to several possible promising developments. First, the same kind of study could be performed for chain of nanoparticles \textcolor{black}{(a first study in the case of a dielectric chain is done in Ref.~\onlinecite{NewPaper})}, for which unexpected effects related to many-body effects~\cite{Add1,Latella1,Latella2} as well as the geometry of the chain could be unveiled. Finally, the same analysis could be performed by going beyond the \textcolor{black}{dipolar} approximation and by including the radiation emitted by the substrate in the energy exchange.

\textcolor{black}{During the review process we became aware of a paper addressing the role of surface waves in the energy transport through a chain of nanoparticles placed in proximity of a planar interface~\cite{NewPaper}. In this work, the authors study both dielectric and metallic nanoparticles, limiting their analysis to a description in terms of electric dipole. With respect to this work, we present a deeper analysis of the dependence of the energy-transport amplification on the chain-surface distance, account for the modified long-distance power-law behavior of the conductance, and describe the major role played by a graphene sheet in the case of metallic nanoparticles.}


\begin{thebibliography}{99}
\bibitem{Rytov}S.~M. Rytov, Y.~A. Kravtsov, V.~I. Tatarskii, \emph{Principles of Statistical Radiophysics}, Vol. 3 (Springer, New York, 1989).
\bibitem{PoldervH}D. Polder and M. van Hove, Phys. Rev. B \textbf{4}, 3303 (1971).
\bibitem{LoomisPRB94}J. J. Loomis and H. J. Maris, Phys. Rev. B \textbf{50}, 18517 (1994).
\bibitem{PendryJPhysCondensMatter99}J. B. Pendry, J. Phys. Condens. Matter \textbf{11}, 6621 (1999).
\bibitem{VolokitinPRB01}A. I. Volokitin and B. N. J. Persson, Phys. Rev. B \textbf{63}, 205404 (2001).
\bibitem{VolokitinPRB04}A. I. Volokitin and B. N. J. Persson, Phys. Rev. B \textbf{69}, 045417 (2004).
\bibitem{JoulainSurfSciRep05}K. Joulain, J.-P. Mulet, F. Marquier, R. Carminati, and J.-J. Greffet, Surf. Sci. Rep. \textbf{57}, 59 (2005).
\bibitem{VolokitinRevModPhys07}A. I. Volokitin and B. N. J. Persson, Rev. Mod. Phys. \textbf{79}, 1291 (2007).
\bibitem{KittelPRL05}A. Kittel, W. M\"{u}ller-Hirsch, J. Parisi, S.-A. Biehs, D. Reddig, and M. Holthaus, Phys. Rev. Lett. \textbf{95}, 224301 (2005).
\bibitem{HuApplPhysLett08}L. Hu, A. Narayanaswamy, X. Chen, and G. Chen, Appl. Phys. Lett. \textbf{92}, 133106 (2008).
\bibitem{NarayanaswamyPRB08}A. Narayanaswamy, S. Shen, and G. Chen, Phys. Rev. B \textbf{78}, 115303 (2008).
\bibitem{RousseauNaturePhoton09}E. Rousseau, A. Siria, G. Joudran, S. Volz, F. Comin, J. Chevrier, and J.-J. Greffet, Nature Photon. \textbf{3}, 514 (2009).
\bibitem{ShenNanoLetters09}S. Shen, A. Narayanaswamy, and G. Chen, Nano Letters \textbf{9}, 2909 (2009).
\bibitem{KralikRevSciInstrum11}T. Kralik, P. Hanzelka, V. Musilova, A. Srnka, and M. Zobac, Rev. Sci. Instrum. \textbf{82}, 055106 (2011).
\bibitem{OttensPRL11}R. S. Ottens, V. Quetschke, S. Wise, A. A. Alemi, R. Lundock, G. Mueller, D. H. Reitze, D. B. Tanner, and B. F. Whiting, Phys. Rev. Lett. \textbf{107}, 014301 (2011).
\bibitem{vanZwolPRL12a}P. J. van Zwol, L. Ranno, and J. Chevrier, Phys. Rev. Lett. \textbf{108}, 234301 (2012).
\bibitem{vanZwolPRL12b}P. J. van Zwol, S. Thiele, C. Berger, W. A. de Heer, and J. Chevrier, Phys. Rev. Lett. \textbf{109}, 264301 (2012).
\bibitem{KralikPRL12}T. Kralik, P. Hanzelka, M. Zobac, V. Musilova, T. Fort, and M. Horak, Phys. Rev. Lett. \textbf{109}, 224302 (2012).
\bibitem{KimNature15}K. Kim \emph{et al.}, Nature \textbf{528}, 387 (2015).
\bibitem{SongNatureNano15}B. Song \emph{et al.}, Nature Nanotechnology \textbf{10}, 253 (2015).
\bibitem{StGelaisNatureNano16}R. St-Gelais, L. Zhu, S. Fan, and M. Lipson, Nature Nanotechnology \textbf{11}, 515 (2016).
\bibitem{KloppstecharXiv}K. Kloppstech \emph{et al.}, Nat. Commun. \textbf{8}, 14475 (2017).
\bibitem{WatjenAPL16}J. I. Watjen, B. Zhao, and Z. M. Zhang, Appl. Phys. Lett. \textbf{109}, 203112 (2016).
\bibitem{PBA-APL2006}P. Ben-Abdallah, Appl. Phys. Lett. \textbf{89}, 113117 (2006).
\bibitem{PBA-PRB2008}P. Ben-Abdallah, K. Joulain, J. Drevillon and C. Le Goff, Phys. Rev. B \textbf{77}, 075417 (2008).
\textcolor{black}{\bibitem{POC2008}P.-O. Chapuis, M. Laroche, S. Volz, and J.-J. Greffet, Appl. Phys. Lett. \textbf{92}, 20 (2008).}
\bibitem{PBA-PRL2011}P. Ben-Abdallah, S.-A. Biehs, and K. Joulain, Phys. Rev. Lett. \textbf{107}, 114301 (2011). 
\bibitem{TschikinEurPhysJB12}M. Tschikin, S.-A. Biehs, F. S. S. Rosa and P. Ben-Abdallah, Eur. Phys. J. B \textbf{85}, 233, (2012).
\bibitem{YannopapasPRL13}V. Yannopapas and N. V. Vitanov, Phys. Rev. Lett. \textbf{110}, 044302 (2013).
\bibitem{YannopapasJPhysChemC13}V. Yannopapas, J. Phys. Chem. C \textbf{117}, 14183 (2013).
\bibitem{BiehsAgarwal2013} S.-A. Biehs and G. S. Agarwal, J. Opt. Soc. Am. B {\bf 30}, 700 (2013).
\bibitem{BaffouOptExpress09}G. Baffou, M. P. Kreuzer, F. Kulzer, and R. Quidant, Opt. Express \textbf{17}, 3291 (2009).
\bibitem{BaffouPRB10}G. Baffou, R. Quidant, and C. Girard, Phys. Rev. B \textbf{82}, 165424 (2010).
\bibitem{BaffouLaserPhotonicsRev13}G. Baffou and R. Quidant, Laser and Photonics Rev. \textbf{7}, 171 (2013).
\bibitem{Add1}P. Ben-Abdallah, R. Messina, S.-A. Biehs, M. Tschikin, K. Joulain, and C. Henkel, Phys. Rev. Lett. \textbf{111}, 174301 (2013).
\bibitem{Add2}M. Nikbakht, J. Appl. Phys. \textbf{116}, 094307 (2014).
\bibitem{Add3}M. Langlais, J.-P. Hugonin, M. Besbes, and P. Ben-Abdallah, Opt. Express \textbf{22}, A577 (2014).
\bibitem{Add4}R. Incardone, T. Emig, and M. Kr\"{u}ger, Europhys. Lett. \textbf{106}, 41001 (2014).
\bibitem{Add5}M. Nikbakht, Europhys. Lett. \textbf{110}, 14004 (2015).
\bibitem{Add6}Y. Wand and J. Wu, AIP Advances \textbf{6}, 025104 (2016).
\bibitem{Add7}O.~R. Choubdar and M. Nikbakht, J. Appl. Phys. \textbf{120}, 144303 (2016).
\bibitem{Add8}J. Dong, J. Zhao, and L. Liu, Phys. Rev. B \textbf{95}, 125411 (2017).
\bibitem{Add9}E. Tervo, Z. Zhang, and B. Cola, Phys. Rev. Materials \textbf{1}, 015201 (2017).
\bibitem{Add10}J. Dong, J.~M. Zhao, and L.~H. Liu, J. Quant. Spectrosc. Radiat. Transf. \textbf{197}, 114 (2017).
\bibitem{Add11}V. Ameri and M. Eghbali-Arani, Eur. Phys. J. D \textbf{71}, 309 (2017).
\textcolor{black}{\bibitem{Add12}K. Asheichyk, B. M\"{u}ller, and M. Kr\"{u}ger Phys. Rev. B 96, 155402 (2017).}
\bibitem{BH}C. F. Bohren and D. R. Huffman, {\it Absorption and Scattering of Light by Small Particles} (Wiley, New York, 1983).
\bibitem{CarminatiOptCommun06}R. Carminati, J.-J. Greffet, C. Henkel, and J.~M. Vigoureux, Opt. Commun. \textbf{261}, 368 (2006).
\bibitem{AlbaladejoOptExpress10}S. Albaladejo, R.~G\'{o}mez-Medina, L.~S. Froufe-P\'{e}rez, H. Marinchio, R. Carminati, J.~F. Torrado, G. Armelles, A. Garc\'{\i}a-Mart\'{\i}n, and J.~J. S\'{a}enz, Opt. Express \textbf{18}, 3556 (2010).
\bibitem{ManjavacasPRB12}A. Manjavacas and F. J. Garc\'{\i}a de Abajo, Phys. Rev. B \textbf{86}, 075466 (2012).
\bibitem{MessinaPRB13}R. Messina, M. Tschikin, S.-A. Biehs, and P. Ben-Abdallah, Phys. Rev. B \textbf{88}, 104307 (2013).
\bibitem{Novotny-book} L. Novotny and B. Hecht, \emph{Principles of Nano-optics} (Cambridge University Press, Cambridge, 2012).
\bibitem{Palik}\textit{Handbook of Optical Constants of Solids}, edited by E. Palik (Academic Press, New York, 1998).
\textcolor{black}{\bibitem{Gold}H. H\"{o}vel, S. Fritz, A. Hilger, U. Kreibig, and M. Vollmer, Phys. Rev. B \textbf{48}, 18178 (1993).}
\bibitem{Persson}B.~N.~J. Persson, and H. Ueba, J. Phys. Condens. Matter \textbf{22}, 462201 (2010).
\bibitem{Volokitin}A.~I. Volokitin and B.~N.~J. Persson, Phys. Rev. B \textbf{83}, 241407(R) (2011).
\bibitem{Svetovoy1}V.~B. Svetovoy, P.~J. van Zwol, and J. Chevrier, Phys. Rev. B \textbf{85}, 155418 (2012).
\bibitem{Ilic1}O. Ilic, M. Jablan, J.~D. Joannopoulos, I. Celanovic, H. Buljan, and M. Solja\v{c}i\'{c}, Phys. Rev. B \textbf{85}, 155422 (2012).
\bibitem{Ilic2}O. Ilic, M. Jablan, J.~D. Joannopoulos, I. Celanovic, H. Buljan, and M. Solja\v{c}i\'{c}, Opt. Express \textbf{20}, A366 (2012)
\bibitem{Messina1}R. Messina, J. P. Hugonin, J.-J. Greffet, F. Marquier, Y. De Wilde, A. Belarouci, L. Frechette, Y. Cordier, and P. Ben-Abdallah, Phys. Rev. B \textbf{87}, 085421 (2013).
\bibitem{Messina2}R. Messina and P. Ben-Abdallah, Sci. Rep. \textbf{3}, 1383 (2013).
\bibitem{Lim2}M. Lim, S.~S. Lee, and B.~J. Lee, Opt. Express \textbf{21}, 22173 (2013).
\bibitem{Phan}A.~D. Phan, S. Shen, and L.~M. Woods, J. Phys. Chem. Lett. \textbf{4}, 4196 (2013).
\bibitem{Liu3}X.~L. Liu and Z. Zhang, Appl. Phys. Lett. \textbf{104}, 251911 (2014).
\bibitem{Liu1}X. Liu, R.~Z. Zhang, and Z. Zhang, ACS Photonics \textbf{1}, 785 (2014).
\bibitem{Svetovoy2}V.~B. Svetovoy and G. Palasantzas, Phys. Rev. Appl. \textbf{2}, 034006 (2014).
\bibitem{Drosdoff}D. Drosdoff, A.~D. Phan, and L.~M. Woods, Advanced Optical Materials \textbf{2}, 1038 (2014).
\bibitem{Zhang1}R.~Z. Zhang, X. Liu, and Z.~M. Zhang, AIP Advances \textbf{5}, 053501 (2015).
\bibitem{Lim1}M. Lim, S. Jin, S.~S. Lee, and B.~J. Lee, Opt. Express \textbf{23}, A240 (2015).
\bibitem{Chang}J.-Y. Chang, Y. Yang, and L. Wang, J. Quant. Spectrosc. Radiat. Transf. \textbf{184}, 58 (2016).
\bibitem{Song}J. Song and Q. Cheng, Phys. Rev. B \textbf{94}, 125419 (2016).
\bibitem{Yin}G. Yin, J. Yang, and Y. Ma, Appl. Phys. Express \textbf{9}, 122001 (2016).
\bibitem{Zheng}Z. Zheng, X. Liu, A. Wang, and Y. Xuan, Int. J. Heat Mass Transfer \textbf{109}, 63 (2017).
\bibitem{Zhao1}B. Zhao and Z.~M. Zhang, ASME J. Heat Transfer \textbf{139}, 022701 (2017).
\bibitem{Simchi}H. Simchi, J. Appl. Phys. \textbf{121}, 094301 (2017).
\bibitem{Lim3}M. Lim, S.~S. Lee, and B.~J. Lee, J. Quant. Spectrosc. Radiat. Transf. \textbf{197}, 84 (2017).
\bibitem{Zhao2}Q. Zhao, T. Zhou, T. Wang, W. Liu, J. Liu, T. Yu, Q. Liao, and N. Liu, J. Phys. D: Appl. Phys. \textbf{50}, 145101 (2017).
\bibitem{Shi}K. Shi, F. Bao, and S. He, ACS Photonics \textbf{4}, 971 (2017).
\bibitem{BenAbdallahAPL15}P. Ben-Abdallah, A. Belarouci, L. Frechette, and S.-A. Biehs, Appl. Phys. Lett. \textbf{107}, 053109 (2015).
\bibitem{Messina3}R. Messina, P. Ben-Abdallah, B. Guizal, and M. Antezza, Phys. Rev. B \textbf{96}, 045402 (2017).
\bibitem{Zhao}B. Zhao, B. Guizal, Z.~M. Zhang, S. Fan, and M. Antezza, Phys. Rev. B \textbf{95}, 245437 (2017).
\bibitem{Geim1}A.~K. Geim and S.~K. Novoselov, Nat. Mater. \textbf{6}, 183 (2007).
\bibitem{Geim2}A.~K. Geim, Science \textbf{324}, 1530 (2009).
\bibitem{BiehsAPL13} S.A.  Biehs, G. S. Agarwal, Appl. Phys. Lett. \textbf{103}, 243112 (2013).
\textcolor{black}{\bibitem{FalkovskyJPhysConfSer08}L. A. Falkovsky, J. Phys. Conf. Ser. \textbf{129}, 012004 (2008).
\bibitem{JablanPRB09}M. Jablan, H. Buljan, and M. Solja\v{c}i\'{c}, Phys. Rev. B \textbf{80}, 245435 (2009).
\bibitem{NewPaper}J. Dong, J. Zhao, and L. Liu, Phys. Rev. B \textbf{97}, 075422 (2018).}
\bibitem{Latella1}I. Latella, P. Ben-Abdallah, S.-A. Biehs, M. Antezza, and R. Messina, Phys. Rev. B \textbf{95}, 205404 (2017).
\bibitem{Latella2}I. Latella, S.-A. Biehs, R. Messina, A.~W. Rodriguez, and P. Ben-Abdallah, Phys. Rev. B \textbf{97}, 035423 (2018).
\end{thebibliography}
\end{document}